\documentclass[12pt]{article}
\usepackage{color}
\usepackage{amssymb}
\usepackage{graphicx}
\usepackage{epstopdf}
\usepackage[hang,small,bf]{caption}
\usepackage{hyperref}

\newcommand{\be}[3]{\begin{equation}  \label{#1#2#3}}
\newcommand{\ee}{\end{equation}}
\newcommand{\ba}{\begin{array}}
\newcommand{\ea}{\end{array}}
\newcommand{\bea}[3]{\begin{eqnarray}  \label{#1#2#3}}
\newcommand{\eea}{\end{eqnarray}}

\newcommand{\nn}{\nonumber}

\let\Large=\large
\let\large=\normalsize

\newcommand{\haken}{\mathbin{\hbox to 8pt{%
                 \vrule height0.4pt width7pt depth0pt
                 \kern-.4pt
                 \vrule height4pt width0.4pt depth0pt\hss}}}

\renewcommand{\theequation}{\thesection.\arabic{equation}}

\def\openone{\leavevmode\hbox{\small1\kern-3.8pt\normalsize1}}
\def\d{\delta}
\def\e{\epsilon}

\def\g{\gamma}

\def\m{\mu}

\def\cl{{\mathcal L}}

\def\bo{{\raise.15ex\hbox{\large$\Box$}}}               % D'Alembertian
\def\face{{\raise.2ex\hbox{$\displaystyle \bigodot$}\mskip-2.2mu \llap {$\ddot
        \smile$}}}                                      % happy face
\def\dg{\dagger}                                     % hermitian conjugate
\def\simlt{\stackrel{<}{{}_\sim}}
\def\simgt{\stackrel{>}{{}_\sim}}

\def\Bar#1{\overline{#1}}                       % big bar
\def\leftrightarrowfill{$\mathsurround=0pt \mathord\leftarrow \mkern-6mu
        \cleaders\hbox{$\mkern-2mu \mathord- \mkern-2mu$}\hfill
        \mkern-6mu \mathord\rightarrow$}       % <--> double differential
\def\dvec#1{\vbox{\ialign{##\crcr
        \leftrightarrowfill\crcr\noalign{\kern-1pt\nointerlineskip}
        $\hfil\displaystyle{#1}\hfil$\crcr}}}           % <--> accent
\def\beq{\begin{equation}}
\def\eeq{\end{equation}}

\def\beqx{\begin{displaymath}}
\def\eeqx{\end{displaymath}}

\def\beqa{\begin{eqnarray}}
\def\eeqa{\end{eqnarray}}

\begin{document}
\DeclareGraphicsExtensions{.jpg,.pdf,.mps,.png}
\begin{flushright}
\baselineskip=12pt
ANL-HEP-PR-09-31\\
EFI-09-02 \\
MADPH-09-1530
\end{flushright}

\begin{center}
\vglue 1.5cm

{\Large\bf $b\rightarrow s$ Transitions in Family-dependent $U(1)^\prime$ Models} \vglue 2.0cm {\Large Vernon Barger$^{a}$, Lisa L. Everett$^{a}$, Jing Jiang$^{a}$, Paul Langacker$^{b}$, \\ Tao Liu$^{c}$ and Carlos E.M.
Wagner$^{c,d,e}$}
\vglue 1cm {
$^a$Department of Physics, University of Wisconsin, Madison, WI 53706\\ \vglue 0.2cm
$^b$School of Natural Science, Institute for Advanced Study, \\
             Einstein Drive, Princeton, NJ 08540
\\ \vglue 0.2cm
$^c$Enrico Fermi Institute and
$^d$Kavli Institute for Cosmological Physics, \\ 
University of Chicago, 5640 S. Ellis Ave., Chicago, IL
60637\\\vglue 0.2cm
$^e$HEP Division, Argonne National Laboratory,
9700 Cass Ave., Argonne, IL 60439
}
\end{center}

\vglue 1.0cm
\begin{abstract}

We analyze flavor-changing-neutral-current (FCNC) effects in the $b\to s$ transitions that are induced by family non-universal $U(1)'$ gauge symmetries. After systematically developing the necessary formalism, we present a correlated analysis for the $\Delta B =1, 2$ processes. %via $b\to s$ transitions. 
We adopt a model-independent approach in which we only require family-universal charges for the first and second generations and small fermion mixing angles. We analyze the constraints on the resulting parameter space from $B_s - \bar B_s$ mixing and the time-dependent  CP asymmetries of the penguin-dominated $B_d \to (\pi, \phi, \eta', \rho, \omega, f_0)K_S$ decays.   Our results indicate that the currently observed discrepancies in some of these modes with respect to the Standard Model predictions can be consistently accommodated within this general class of models.

\end{abstract}

\newpage

\section{Introduction}

%During the past several decades, great progress has been made in understanding $CP$-violating phenomena in the meson systems. With the advent of

The origin of $CP$ violation, which was first observed in the kaon system  four decades ago \cite{cronin}, has remained one of the fundamental questions of elementary particle physics.  In recent years, the B factories have established that the Standard Model (SM) picture of $CP$ violation, in which all $CP$-violating effects are generated by the single phase $\delta_{\rm CKM}$ in the Cabibbo-Kobayashi-Maskawa (CKM) quark mixing matrix  \cite{1963, 1973}, is consistent with the observed pattern of $CP$-violating phenomena in both the $B_d$ and $K$ meson systems~\cite{PDG}.  However, as the SM cannot account for the baryon asymmetry in the Universe today\cite{baryonasymmetry}, new physics (NP) is {\it necessarily} required to describe all observed phenomena with $CP$-violation involved.   

One arena to seek the NP effects is in flavor-changing neutral current transitions (FCNC) where the SM contributions first appear at the one-loop level and the NP effects can be competitive. 
%Even though the $K$ and $B_d$ systems provide stringent constraints on many of the flavor-changing and CP-violating couplings that can generically be present in the NP theories, .  
The emblematic set of such processes is the set of $b\to s$ transitions, which include $B_s-\bar B_s$ mixing and the set of neutral $B_d$ meson decays which occur via $b \to s \bar qq$ $(q = u, d, c, s)$ transitions.  Several of these processes are also of interest because recent measurements exhibit discrepancies with the SM predictions at the level of a few standard deviations, which may suggest the intriguing possibility of physics beyond the SM.  The current status of the data is as follows:\\

\noindent $\bullet$  {\bf $B_s-\bar B_s$ mixing phase}.  The standard way to parametrize NP in $B_s-\bar B_s$ mixing is to express the off-diagonal mixing matrix element as follows:
\begin{eqnarray}
M_{12}^{B_s}=(M_{12}^{B_s})_{\rm SM} C_{B_s} e^{2 i \phi_{B_s}^{\rm NP}}. \label{001}
\end{eqnarray}
The SM predicts that $C_{B_s}=1$ and $\phi_{B_s}^{\rm NP}=0$.  Though the data indicate that $C_{B_s}$ does not differ significantly from unity, the results of a recent analysis~\cite{Bona:2008jn} suggest that $\phi_{B_s}^{\rm NP}$ deviates from zero at the $3 \sigma$ level (see Table~\ref{table1}). This analysis combines all the available experimental results on $B_s$  mixing, including the new tagged analyses of $B_s \to \psi \phi$ by CDF~\cite{Aaltonen:2007he} and D$\emptyset$~\cite{:2008fj} (note that no single measurement yet has a $3\sigma$ significance.). The discrepancy disfavors NP scenarios which obey minimal flavor violation (MFV), {\it i.e.}, with $\phi_{B_s}^{\rm NP}\approx 0$, and instead suggests NP which exhibits flavor violation in the $b\rightarrow s$ transitions (e.g., see~\cite{Tarantino:2008pb} and references therein). For convenience, in Table~\ref{table1} we also give the data in terms of $A_s^{\rm NP}/A_s^{\rm SM}$ and $\phi_s^{\rm NP}$ which are related to $C_{B_s}$ and $\phi_{B_s}^{\rm NP}$ according to  
\begin{eqnarray}
C_{B_s} e^{2i\phi_{B_s}^{\rm NP}}  = 1 +  \frac{A_s^{\rm NP}}{A_s^{\rm SM}}e^{2 i \phi_s^{\rm NP}}. \label{002} 
\end{eqnarray}\\

\begin{table}[th]
\begin{center}
\begin{tabular}{@{}ccc}
Observable  & $1 \sigma$ C.L.& $2 \sigma$ C.L.  \\
\hline
\hline
$\phi_{B_s}^{\rm NP} [^\circ]$  (S1)            & -20.3 $\pm$ 5.3 & [-30.5,-9.9] \\
    $\phi_{B_s}^{\rm NP} [^\circ]$                         (S2)       & -68.0 $\pm$ 4.8 & [-77.8,-58.2] \\
$C_{B_s}$                           & 1.00 $\pm$ 0.20 & [0.68,1.51] \\
\hline
$\phi_s^{\rm NP} [^\circ]$  (S1)            & -56.3 $\pm$ 8.3 & [-69.8,-36.0] \\
$A_s^{\rm NP}/A_s^{\rm SM}$ (S1)     & 0.66 $\pm$ 0.28 & [0.24,1.11] \\
    $\phi_s^{\rm NP} [^\circ]$         (S2)       & -79.1 $\pm$ 2.6 & [-84.0,-72.8] \\
  $A_s^{\rm NP}/A_s^{\rm SM}$ (S2)     &1.78 $\pm$ 0.03& [1.53,2.19] \\
\hline
\hline
\end{tabular}
\end{center}
\caption {Fit results for the $B_s - \bar B_s$ mixing parameters~\cite{Bona:2008jn}. The two $\phi_{B_s}^{\rm NP}$ solutions (``S1'' and ``S2'') result from measurement ambiguities; see~\cite{Bona:2008jn} for details.}
\label{table1}
\end{table}

\noindent $\bullet$ {\bf $CP$ asymmetries in neutral $B_d$ decays}.  The set of neutral $B_d$ decays in question is the set of QCD penguin-dominated charmless decays that occur via $b \to s \bar qq$ $(q = u, d, c, s)$ transitions.  The $CP$ asymmetries of such decays into a final $CP$-eigenstate $f_{CP}$ are given by
\begin{eqnarray}
{\mathcal A}_{f_{CP}}(t)
&=& \frac{\Gamma(\bar{B}_d(t)\to f_{CP})-\Gamma(B_d(t)\to f_{CP})}
         {\Gamma(\bar{B}_d(t)\to f_{CP})+\Gamma(B_d(t)\to f_{CP})}\Big|_{\Delta \Gamma_{B_d}=0}
\nonumber \\
&=& -{\mathcal C}_{f_{CP}}\cos (\Delta M_{B_d}t) + {\mathcal S}_{f_{CP}} \sin (\Delta M_{B_d}t),
\end{eqnarray}
in which ${\mathcal C}_{f_{CP}}$ and ${\mathcal S}_{f_{CP}}$ are direct and mixing-induced $CP$ asymmetry parameters.  The SM predictions for many decays of this type, including $B_d\to \psi K_S$ and $B_d \to (\phi, \eta', \pi, \rho, \omega, f_0) K_S$ are as follows: 
%(e.g., see~\cite{Buchalla:1995vs}) 
\begin{eqnarray}
-\eta_{f_{CP}}{\mathcal S}_{f_{CP}} = \sin 2 \beta + {\mathcal O} (\lambda^2),  \ \ \ \  {\mathcal C}_{f_{CP}} = 0 + {\mathcal O} (\lambda^2),  \label{TDCP}
\end{eqnarray}
with $\beta \equiv \arg\left[-(V_{cd}V_{cb}^*)/(V_{td}V_{tb}^*)\right]$,  $\lambda=\sin\theta_c$ being the Cabibbo angle, and $\eta_{f_{CP}}=\pm1$ being the $CP$ eigenvalue for the final state $f_{CP}$.  However, the central values of $\sin 2 \beta$ directly measured from the penguin-dominated modes are systematically below the SM prediction and the results obtained from measuring the charmed $B_d \to \psi K_S$ mode. Meanwhile, the central values of the direct $CP$ asymmetry measured from $B_d\to \phi K_S$ and $B_d \to \omega K_S$ modes are also small  compared to that obtained from the $B_d \to \psi K_S$ mode (see Table~\ref{table2}).  Given that the $B_d\to \psi K_S$ decay is dominated by tree-level amplitude in the SM, large absolute values for $\Delta {\mathcal S}_{f_{CP}}=-\eta_{f_{CP}}{\mathcal S}_{f_{CP}}+ \eta_{\psi K_S}{\mathcal S}_{\psi K_S}$ and $\Delta {\mathcal C}_{f_{CP}}={\mathcal C}_{f_{CP}} - {\mathcal C}_{\psi K_S}$ may imply interesting NP in the $b\to s$ transitions. .
\begin{table}[t]
%\vspace{0.4cm}
\begin{center}
\begin{tabular}{|c|c|c|}
  \hline
$f_{CP}$ & $-\eta_{CP} {\mathcal S}_{f_{CP}}$ (1$\sigma$ C.L.) & ${\mathcal C}_{f_{CP}}$(1$\sigma$ C.L.)   \\  \hline
$\psi K_S$  & $+0.672\pm0.024 $ & $+0.005\pm0.019 $   \\ \hline 
$\phi K_S$ & $+0.44^{+0.17}_{-0.18} $ & $-0.23\pm0.15 $  \\
$\eta^\prime K_S$ & $+0.59\pm0.07  $ & $-0.05\pm0.05$  \\  
$\pi K_S$ & $+0.57\pm0.17$ & $+0.01\pm0.10$  \\
$\rho K_S$ & $+0.63^{+0.17}_{-0.21}$ & $-0.01\pm0.20 $  \\ 
$\omega K_S$ & $+0.45\pm0.24$ & $-0.32\pm0.17 $  \\ 
$f_0 K_S$ & $+0.62^{+0.11}_{-0.13}$ & $0.10\pm0.13$ \\ \hline
\end{tabular}
\end{center}
\caption{World averages of the experimental results for the $CP$
  asymmetries in $B_d$ decays via $b\to\bar qq s$ transitions~\cite{Barberio:2008fa}.} \label{table2}
\end{table}

To account for these discrepancies appearing in $B_s - \bar B_s$ mixing and $B_d$ decays, a number of NP scenarios have been studied, including low energy supersymmetry and models with warped extra dimensions, among others~\cite{NPmodels}.   In many of these scenarios, the effects of NP in the $b\to s$ transitions are loop-suppressed and can compete with SM contributions.  The most popular and well-studied scenarios are models with minimal flavor violation (MFV), in which the only source of $CP$ violation is the single irremovable phase of the Cabibbo-Kobayashi-Maskawa mixing matrix.   MFV scenarios, however, face difficulties in that they do not generally allow for a nonvanishing $\phi_{B_s}$. %MORE DETAIL HERE.

In this work, we will study the constraints from $b\to s$ transitions on models with family non-universal gauged $U(1)'$ symmetries.  Additional $U(1)^\prime$ gauge symmetries are present in many well-motivated extensions of the SM, such as grand unified and/or string models (e.g., see~\cite{Langacker:2008yv} for a review).   Such scenarios are of particular interest because unlike the scenarios studied above, they allow for the intriguing possibility of {\it tree-level} FCNC, with contributions that are competitive with the SM even for small $U(1)^\prime$ couplings.   Depending on the details of the model, family-dependent $U(1)^\prime$ scenarios can result in new FC operators and/or modified Wilson coefficients to the existing SM operators in the operator product expansion, providing a rich framework beyond MFV to explore FCNC and $CP$-violating effects.

We follow the general framework for addressing $Z'$-induced FCNC as developed in~\cite{Langacker:2000ju} and systemize its application to $b\to s$ transitions.  Rather than considering specific $U(1)^\prime$ models, we adopt a model-independent approach in which the main restrictions are family universal charges for the first and second generations and small fermion mixing angles.  We also neglect the effects of $Z-Z'$ mixing (which are known to be small), and assume the absence of any exotic fermions that could mix with the usual SM fermions through non-universal $Z'$ couplings, which may also result in nontrivial FCNC effects (e.g., see~\cite{Langacker:2008yv}). 

This work is an extension of our earlier work~\cite{Barger:2009eq,Barger:2004qc}, in which we performed a {\it correlated} analysis of the $\Delta B =1, 2$ processes mentioned above for a specific set of $U(1)^\prime$ scenarios.  That analysis was in contrast to other studies of $U(1)^\prime$ scenarios based on mode-by-mode analyses~\cite{Barger:2003hg}.   The purpose of this paper is twofold: first, to provide more details of the formalism and analysis than were given explicitly in~\cite{Barger:2009eq}, and second, to analyze a more general set of $U(1)^\prime$ models.  Our results demonstrate that the $b\to s$ transitions not only place important constraints on family non-universal $Z'$ couplings and mass scale, but also that family non-universal $U(1)^\prime$ scenarios can explain the currently observed discrepancies with the SM predictions for $B_s - \bar B_s$ mixing and the time-dependent  $CP$ asymmetries of the penguin-dominated $B_d \to (\pi, \phi, \eta', \rho, \omega, f_0)K_S$ decays.

This paper is structured as follows.  We begin by providing an overview of the formalism of the $Z'$ induced FCNC effects in the $b\to s$ transitions and present the effective Hamiltonian for the processes of interest at the $b$ quark mass scale in Section 2.   In Section 3, first we analyze the FCNC constraints within several special limits of the general $U(1)^\prime$ parameter space, and then turn to a more general analysis.  Our summary and conclusions are presented in Section 4. 

%%%%%%%%%%%%%%%%%%%%%%%%%%%%%%%%%%%%%%%%%
\section{Theoretical Background}
\subsection{Formalism of $Z'$-induced FCNC Effects}
\label{section 2}
%%%%%%%%%%%%%%%%%%%%%%%%%%%%%%%%%%%%%%%%%

The general framework for studying $Z'$-induced FCNC Effects has been developed in~\cite{Langacker:2000ju}. In this section, we will systematically formalize its applications to the case of the $b\to s$ transitions (the generalization to $b\to d$ transitions is straightforward).

We begin by considering the SM extended by a single additional $U(1)^\prime$ gauge symmetry (the generalization to multiple $U(1)^\prime$ gauge symmetries is straightforward).  In this theory, the neutral current Lagrangian in the SM gauge eigenstate basis is given by
\begin{eqnarray}
  \cl_{NC}=-eJ^{\m}_{\mbox{\tiny em}}A_{\m} - g_1 J_Z^\mu Z_\mu
           - g_2 J_{Z'}^\mu Z'_\mu,
\end{eqnarray}
in which $A_\mu$ is the $U(1)_{\mbox{\tiny em}}$ gauge boson, $Z_\mu$ is the massive electroweak (EW) neutral gauge boson, $Z'_\mu$ is the gauge boson associated with the additional Abelian
gauge symmetry, and  $g_1=g/\cos\theta_W$ and $g_2$ are the gauge couplings of the $Z_\mu$ and $Z'_\mu$ bosons, respectively. The currents are given by
\begin{eqnarray}
  J_Z^\mu&=&\sum_{\psi} \sum\limits_i \Bar{\psi}_i \g^{\mu} 
    \left[ \e^{\psi_L}_i P_L + \e^{\psi_R}_iP_R\right]\psi_i, \label{J1}\\[1ex]
  J_{Z'}^\mu&=&\sum_{\psi}\sum\limits_{i,j} \Bar{\psi}_i \g^{\mu} 
    \left[\tilde \e^{\psi_L}_{ij} P_L 
     + \tilde \e^{\psi_R}_{ij}P_R\right]\psi_j,\label{J2}
\end{eqnarray}
in which $\psi$ labels the SM fermions, $i$ and $j$ are family indices, and $P_{R,L}=(1\pm\g_5)/2$. The (family universal) SM chiral charges are given by
\begin{eqnarray}
 \e^{\psi_L}_i=t_3^{\psi_L}-\sin^2\theta_WQ_{\psi_L}, \ \ \ \  \e^{\psi_R}_i=-\sin^2\theta_W Q_{\psi_R}, \label{150}
\end{eqnarray}
in which $t_3^{\psi_L}$ denotes the third component of the weak isospin
and $Q_{\psi_{L,R}}$ are the electric charges of  $\psi_{L,R}$.   Without loss of generality, the $Z'$ chiral charges 
can be diagonalized  by choosing the appropriate gauge basis for the fermions:
\begin{eqnarray}
\tilde \e^{\psi_{L,R}}_{ij} = \tilde \e^{\psi_{L,R}}_i \delta_{ij}. \label{151}
\end{eqnarray}
In particular, $SU(2)_L$ symmetry requires that
\begin{eqnarray}
\tilde \e^{u_L}_{i} \equiv \tilde \e^{d_{L}}_{i}, \ \ \ \ \tilde \e^{e_L}_i \equiv \tilde \e^{\nu_L}_i.  \label{152}
\end{eqnarray}
%as a generic requirement of $Z'$ model construction. 
%Flavor changing effects immediately arise if the $\tilde \e$ are non-diagonal matrices. 
If the diagonal $U(1)^\prime$ chiral charges are non-universal, flavor-changing (FC) $Z'$ couplings are generically induced by  fermion mixing. The fermion Yukawa matrices $h_{\psi}$ in the weak eigenstate
basis are diagonalized by the unitary matrices $V_{\psi_{L,R}}$, such that
\begin{eqnarray}
  h_{\psi,diag}=V_{\psi_R} h_{\psi} V_{\psi_L}^{\dg}, \label{153}
\end{eqnarray}
and the CKM matrix is given by
\begin{eqnarray}
  V_{\rm CKM} = 
    V_{u_L} V_{d_L}^{\dg}. \label{154} 
  \label{BCKM}
\end{eqnarray}
Hence, the chiral $Z'$ couplings in the fermion mass eigenstate
basis take the form: 
\begin{eqnarray}
  B^{\psi_L}\equiv
    V_{\psi_L}\tilde \e^{\psi_L} V_{\psi_L}^{\dg}\;,
    \qquad
  B^{\psi_R} \equiv
    V_{\psi_R} \tilde \e^{\psi_R} V_{\psi_R}^{\dg}\;.
  \label{Bij}
\end{eqnarray}
However, it is known that the constraints from $K- \bar K$ mixing and from $\mu-e$ conversion in muonic atoms 
exclude significant non-universal effects for the first two families,
% for a TeV-scale $Z'$ with EW couplings~\cite{Langacker:2000ju}, 
which suggests that
\begin{eqnarray}
B^{\psi_{L,R}} =\begin{pmatrix} {  B^{\psi_{L,R}}_{11} &0&B^{\psi_{L,R}}_{13} \cr 0 &   B^{\psi_{L,R}}_{11}& B^{\psi_{L,R}}_{23} \cr B^{\psi_{L,R}*}_{13} & B^{\psi_{L,R}*}_{23} &  B^{\psi_{L,R}}_{33} }  \end{pmatrix}, \label{160}
\end{eqnarray}
at least for the down-type quarks and $e$, $\mu$, $\tau$ leptons.
%\footnote{Actually, this pattern is only required for down-type quark and leptons. Here we extend it to up-type fermions just for simplicity.}. 
The most straightforward way to achieve this coupling structure is to assume universal $U(1)^\prime$ 
charges for the down-type fermions of the first two families, {\it i.e.}, 
\begin{eqnarray}
\tilde \e^{\psi_{L,R}} =\begin{pmatrix} {\tilde \e^{\psi_{L,R}}_1 &0&0 \cr 0 &  \tilde  \e^{\psi_{L,R}}_1  & 0 \cr 0 & 0& \tilde  \e^{\psi_{L,R}}_3 } \end{pmatrix}. \label{161}
\end{eqnarray}
With the unitary matrices $V_{\psi_{L,R}}$ written as 
\begin{eqnarray}
 V_{\psi_{L,R}} =\begin{pmatrix} { W_{\psi_{L,R}} & X_{\psi_{L,R}} \cr  Y_{\psi_{L,R}} & Z_{\psi_{L,R}} } \end{pmatrix}, \label{162}
\end{eqnarray}
where $W_{\psi_{L,R}}$ is a $2\times 2$ submatrix, one obtains 
\begin{eqnarray}
B^{\psi_{L,R}} &=&\begin{pmatrix} {\tilde \e^{\psi_{L,R}}_1 W_{\psi_{L,R}}^\dagger W_{\psi_{L,R}} + \tilde \e^{\psi_{L,R}}_3 Y_{\psi_{L,R}}^\dagger Y_{\psi_{L,R}}  & \tilde \e^{\psi_{L,R}}_1 W_{\psi_{L,R}}^\dagger X_{\psi_{L,R}} + \tilde \e^{\psi_{L,R}}_3 Y_{\psi_{L,R}}^\dagger Z_{\psi_{L,R}} \cr \tilde \e^{\psi_{L,R}}_1 X_{\psi_{L,R}}^\dagger W_{\psi_{L,R}} + \tilde \e^{\psi_{L,R}}_3 Z_{\psi_{L,R}}^\dagger Y_{\psi_{L,R}} &  \tilde \e^{\psi_{L,R}}_1 X_{\psi_{L,R}}^\dagger X_{\psi_{L,R}} + \tilde \e^{\psi_{L,R}}_3 Z_{\psi_{L,R}}^\dagger Z_{\psi_{L,R}} } \end{pmatrix}.\nonumber \\ \label{163}
\end{eqnarray}
Therefore, in the limit of small fermion mixing angles or small $X_{\psi_{L,R}}$, $Y_{\psi_{L,R}}$ elements, a $Z'$ coupling structure of the type given in Eq.~(\ref{160}) is produced, in which
\begin{eqnarray}
&&  B^{\psi_{L,R}}_{11} = \tilde \e^{\psi_{L,R}}_1, \ \ \ \    B^{\psi_{L,R}}_{33} = \tilde \e^{\psi_{L,R}}_3    \nonumber \\
&& B^{\psi_{L,R}}_{13}, B^{\psi_{L,R}}_{23} \sim {\mathcal O} (X_{\psi_{L,R}},Y_{\psi_{L,R}}), \label{180}
\end{eqnarray}
such that $B^{\psi_{L,R}}_{13}$ and $B^{\psi_{L,R}}_{23}$ are in general both complex parameters.

EW symmetry breaking induces $Z-Z'$ mixing, such that the gauge eigenstates $Z_\mu$ and $Z'_\mu$ are related to the mass eigenstates $Z^{(n)}_\mu$ ($n=1,2$) by an orthogonal transformation.  In the mass eigenstate basis, the Lagrangian couplings are given by 
\begin{eqnarray}
  \cl_{NC}^Z= -  \left[g_1\cos\theta J_Z^\mu 
      + g_2\sin\theta J_{Z'}^\mu\right] Z^{(1)}_{\m}
    - \left[   - g_1\sin\theta J_Z^\mu + g_2\cos\theta J_{Z'}^\mu 
   \right] Z^{(2)}_{\m}, \label{LZ}
\end{eqnarray}
where $\theta$ is the $Z-Z'$ mixing angle, $J_Z^\mu$ is given in Eq.~(\ref{J1}), and $J_{Z'}^\mu$ is of the form of Eq.~(\ref{J2}) with $\tilde \e^{\psi_{L,R}}$ replaced by $B^{\psi_{L,R}}$ from Eq.~(\ref{Bij}).   In this analysis, we neglect kinetic mixing since it simply amounts to a redefinition of the unknown $Z'$ couplings.\footnote{Kinetic mixing allows the redefined $Z'$ charges to have a component of
weak hypercharge, which would otherwise not be allowed. This feature is
irrelevant for the purposes of this paper.}

At the EW scale, the tree-level four-fermion interactions are described by the product of gauge currents
\begin{eqnarray}
  \cl_{eff}&=&{-4G_F\over\sqrt{2}}
    \left(\rho_{eff} {J_Z}^2 
    + 2wJ_Z\cdot J_{Z'}+ y {J_{Z'}}^2\right) \nonumber \\[1ex]
  &=&{-4G_F\over\sqrt{2}}\sum\limits_{\psi,\chi}\sum\limits_{i,j,m,n}
  \left[
     { C}^{ij}_{mn} S^{ij}_{mn}
    + \tilde{ C}^{ij}_{mn} \tilde S^{ij}_{mn}
    + { D}^{ij}_{mn} T^{ij}_{mn}
    + \tilde{ D}^{ij}_{mn} \tilde T^{ij}_{mn}
  \right]. \label{111}
\end{eqnarray}
In Eq.~(\ref{111}), the local current-current operators are\footnote{These operators are not all
independent. For couplings of four fermions of the same type,
$\psi=\chi$, e.g.\ four charged leptons, one has
$S^{ij}_{mn}=S^{mn}_{ij}$, $\tilde S^{ij}_{mn}=\tilde S^{mn}_{ij}$ and
$T^{ij}_{mn}=\tilde T^{mn}_{ij}$.}
 ($i,j,m,n$ are family indices):
 \begin{eqnarray}
  S^{ij}_{mn}
    =\left(\Bar{\psi}_i\g^{\m}P_L\psi_j\right)
    \left(\Bar{\chi}_m\g_{\m}P_L\chi_n\right),&\qquad&
  \tilde S^{ij}_{mn}
    =\left(\Bar{\psi}_i\g^{\m}P_R\psi_j\right)
    \left(\Bar{\chi}_m\g_{\m}P_R\chi_n\right),\label{op}  \nonumber \\[1ex]
  T^{ij}_{mn}
    =\left(\Bar{\psi}_i\g^{\m}P_L\psi_j\right)
    \left(\Bar{\chi}_m\g_{\m}P_R\chi_n\right),&\qquad&
  \tilde T^{ij}_{mn}
    =\left(\Bar{\psi}_i\g^{\m}P_R\psi_j\right)
    \left(\Bar{\chi}_m\g_{\m}P_L\chi_n\right), \label{112}
\end{eqnarray}
and the coefficients are
\begin{eqnarray}
 { C}^{ij}_{mn}&=& \rho_{eff}\d_{ij}\d_{mn}\e^{\psi_L}_i\e^{\chi_L}_m
    + w\d_{ij}\e^{\psi_L}_iB^{\chi_L}_{mn}
    + w\d_{mn}\e^{\chi_L}_mB^{\psi_L}_{ij}
    + yB^{\psi_L}_{ij}B^{\chi_L}_{mn},\nonumber \\[1ex]
  \tilde{ C}^{ij}_{mn}&=& \rho_{eff}\d_{ij}\d_{mn}\e^{\psi_R}_i\e^{\chi_R}_m
    + w\d_{ij}\e^{\psi_R}_iB^{\chi_R}_{mn}
    + w\d_{mn}\e^{\chi_R}_mB^{\psi_R}_{ij}
    + yB^{\psi_R}_{ij}B^{\chi_R}_{mn}, \nonumber \\[1ex]
  { D}^{ij}_{mn}&=& \rho_{eff}\d_{ij}\d_{mn}\e^{\psi_L}_i\e^{\chi_R}_m
    + w\d_{ij}\e^{\psi_L}_i B^{\chi_R}_{mn}
    + w\d_{mn}\e^{\chi_R}_mB^{\psi_L}_{ij}
    + yB^{\psi_L}_{ij}B^{\chi_R}_{mn},\nonumber \\[1ex]
  \tilde{ D}^{ij}_{mn}&=& \rho_{eff}\d_{ij}\d_{mn}\e^{\psi_R}_i\e^{\chi_L}_m
    + w\d_{ij}\e^{\psi_R}_i B^{\chi_L}_{mn}
    + w\d_{mn}\e^{\chi_L}_mB^{\psi_R}_{ij}
    + yB^{\psi_R}_{ij}B^{\chi_L}_{mn}, \label{113}
\end{eqnarray}
in which
\begin{eqnarray}
  \rho_{eff}&=&\rho_1\cos^2\theta + \rho_2\sin^2\theta,
    \qquad \rho_a={M_W^2\over M_a^2\cos^2\theta_W},\nonumber \\[1ex]
  w&=&{g_2\over g_1}\sin\theta\cos\theta(\rho_1-\rho_2),\nonumber \\[1ex]
  y&=&{\left(g_2\over g_1\right)}^2(\rho_1\sin^2\theta 
     + \rho_2\cos^2\theta). \label{114}
\end{eqnarray}
In Eqs.~(\ref{114}), $M_a$ denotes the masses of the neutral gauge boson mass eigenstates, and 
$\theta_W$ is the EW mixing angle. 
We do not specify  the $\psi$ and $\chi$ dependence of the coefficients $C, \tilde C, D, \tilde D$ in Eqs.~(\ref{113}), which can be understood from the context. 
%This effective Hamiltonian describes the interactions between the $Z$ and $Z'$ currents at tree level. 

For $b\to s$ transitions, the local operators are given by $S_{mn}^{bs}$, $\tilde S_{mn}^{bs}$, $T_{mn}^{bs}$ and $\tilde  T_{mn}^{bs}$, with coefficients that are given by 
\begin{eqnarray}
  C^{bs}_{mn}&=&w\d_{mn}\e^{\chi_L}_mB_{bs}^L
    + yB_{bs}^LB^{\chi_L}_{mn}\;, \nonumber \\[1ex]
  \tilde{C}^{bs}_{mn}&=& w\d_{mn}\e^{\chi_R}_mB_{bs}^R
    + yB_{bs}^RB^{\chi_R}_{mn}\;,\nonumber \\[1ex]
  D^{bs}_{mn}&=&  w\d_{mn}\e^{\chi_R}_mB_{bs}^L
    + yB_{bs}^LB^{\chi_R}_{mn}\;,\nonumber \\[1ex]
  \tilde{D}^{bs}_{mn}&=& w\d_{mn}\e^{\chi_R}_mB_{bs}^R
    + yB_{bs}^RB^{\chi_L}_{mn}\;. \label{115}
\end{eqnarray}
With the $Z-Z'$ mixing angle neglected,  the coefficients can be written as 
\begin{eqnarray}
  C^{bs}_{mn}&=& \left(\frac{g_2 M_Z}{g_1 M_{Z'}}\right)^2B_{bs}^LB^{\chi_L}_{mn}\;,\nonumber \\[1ex]
  \tilde{C}^{bs}_{mn}&=&  \left(\frac{g_2 M_Z}{g_1 M_{Z'}}\right)^2 B_{bs}^RB^{\chi_R}_{mn}\;, \nonumber \\[1ex]
  D^{bs}_{mn}&=&   \left(\frac{g_2 M_Z}{g_1 M_{Z'}}\right)^2  B_{bs}^LB^{\chi_R}_{mn}\;,\nonumber \\[1ex]
  \tilde{D}^{bs}_{mn}&=&   \left(\frac{g_2 M_Z}{g_1 M_{Z'}}\right)^2 B_{bs}^RB^{\chi_L}_{mn}. \label{116}
\end{eqnarray}
For convenience, in the following we will resolve the factor $g_2 M_Z/(g_1 M_{Z'})$ into the $B$ elements or the chiral couplings.
% $B$ elements or chiral couplings.

At tree level, there are three classes of $b\to s$ transitions which are sensitive to the possible NP effects that result  from an additional family non-universal $U(1)^\prime$ symmetry:  $b \to s \bar qq$ transitions,  $b \to s \bar ll$ transitions, and $B_s-\bar B_s$ mixing. Here ``q'' and ``l'' denote quarks and leptons, respectively.
For the $b \to s \bar qq$ transitions, the $Z'$ effects are described by the effective Hamiltonian 
\begin{eqnarray}
{\cal H}_{\rm eff}^{Z'} (b \to s \bar q q) 
&=&  \frac{2G_F}{\sqrt{2}}  \Big( (\bar sb)_{V-A} \sum_q (C^{bs}_{qq} (\bar qq)_{V-A} + D^{bs}_{qq} (\bar qq)_{V+A}) \nonumber \\ && +  (\bar sb)_{V+A} \sum_q (\tilde D^{bs}_{qq} (\bar qq)_{V-A} + \tilde C^{bs}_{qq} (\bar qq)_{V+A}) \Big)+ \mbox{h.c.},  \label{118}
\end{eqnarray}
in which the sum is over the active quarks for a given process. These $Z'$-induced FCNC effects can be understood as corrections to the SM operators or to the new penguin operators defined in~\ref{Operators}, since both lead to the same hadronic matrix elements.  Explicitly, comparing Eq.~(\ref{118}) with
\begin{eqnarray}
&&{\cal H}_{\rm eff}^{Z'} (b \to s \bar q q)
%&=& - \frac{G_F}{\sqrt{2}} V_{tb} V_{ts}^* \sum_q \left( \Delta C_3 Q_3^{(q)} +
%  \Delta C_5 Q_5^{(q)}  \right. \left. + \Delta C_7 Q_7^{(q)} + \Delta C_9 Q_9^{(q)} \right) + \mbox{h.c.} 
%\nonumber\\ 
= %- \frac{G_F}{\sqrt{2}} V_{tb} V_{ts}^*\, \times \, 
\nonumber \\
&& - \frac{G_F}{\sqrt{2}} V_{tb} V_{ts}^*
\Big[ ({\bar s}b)_{V-A} 
\sum_q \left( ( \Delta C_3 + \Delta C_9 \frac{3}{2} e_q ) ({\bar q}q)_{V-A}
  + ( \Delta C_5 + \Delta C_7 \frac{3}{2} e_q ) ({\bar q}q)_{V+A} \right) \nonumber \\ &&  +({\bar s}b)_{V+A} 
\sum_q \left( ( \Delta \tilde C_3 + \Delta \tilde C_9 \frac{3}{2} e_q ) ({\bar q}q)_{V+A}
  + ( \Delta \tilde C_5 + \Delta \tilde C_7 \frac{3}{2} e_q ) ({\bar q}q)_{V-A} \right) \Big]+\mbox{h.c.}, \nonumber \\ 
  %&& +\mbox{h.c.}, 
  \label{117}
\end{eqnarray}
results in $4n_q$ equations ($n_q$ is the number of active quarks in the final states):
\begin{eqnarray}
\Delta C_3 + \Delta C_9 \frac{3}{2} e_q &=&\frac{-2}{V_{tb}V_{ts}^*} C_{qq}^{bs},  \nonumber \\
\Delta C_5 + \Delta C_7 \frac{3}{2} e_q &=&\frac{-2}{V_{tb}V_{ts}^*} D_{qq}^{bs},  \nonumber \\
\Delta \tilde C_3 + \Delta \tilde C_9 \frac{3}{2} e_q &=&\frac{-2}{V_{tb}V_{ts}^*} \tilde C_{qq}^{bs},  \nonumber \\
\Delta \tilde C_5 + \Delta \tilde C_7 \frac{3}{2} e_q &=&\frac{-2}{V_{tb}V_{ts}^*} \tilde D_{qq}^{bs},   \label{121}
\end{eqnarray}
where $\Delta C$ denotes $Z'$ correction to the Wilson coefficients of the SM operators and $\Delta \tilde C$ denotes
the Wilson coefficients of the operators beyond the SM ones. For charmless processes with $q$ from the first two families, these equations are solvable because of the following relation obeyed by the down-type quark couplings:
\begin{eqnarray}
B_{11}^{\psi_{L,R}} &=& B_{22}^{\psi_{L,R}}, \label{122}
\end{eqnarray}
which is extracted from Eq.~(\ref{160}). The $Z'$ corrections to the Wilson coefficients are then found to be\footnote{Though the solutions to Eq.~(\ref{121}) are not unique in the case with $n_q=1$, 
the physics is unaffected since it is only sensitive to the linear combinations on the left hand side of Eq.~(\ref{121}). For the charmed processes where generally we have $n_q=1$ , 
the formula in Eq. (\ref{120}) can also be applied as long as Eq.~(\ref{122}) holds for the up-type quarks. If Eq.~(\ref{122}) does not hold, then the ``$uu$'' indices in these formula need to be replaced by ``$cc$''. } 
%\begin{eqnarray}
%B_{11}^{\psi_{L,R}} &=& B_{22}^{\psi_{L,R}},  \  \ \ \ {\rm or} \nonumber \\
%C_{uu}^{bs} = C_{ss}^{bs}, &&  D_{uu}^{bs} = D_{ss}^{bs}, \nonumber \\
%\tilde C_{uu}^{bs} = \tilde C_{ss}^{bs}, &&  \tilde D_{uu}^{bs} = \tilde D_{ss}^{bs},
%\end{eqnarray}
%Note, $\Delta C^q$ and $\Delta \tilde C^q$ are not family-universal, because $C_{qq}^{bs}$, $D_{qq}^{bs}$, 
%$\tilde C_{qq}^{bs}$ and $\tilde D_{qq}^{bs}$ are not family-universal. Without loss of generality, we can take 
%\begin{eqnarray}
%\Delta C^{u} = \Delta C^{d_i}, \ \ \ \  \Delta \tilde C^{u_i} = \Delta \tilde C^{d_i}
%\end{eqnarray} 
%for up- and down-type fermions in the same family. 
\begin{eqnarray}
\Delta C_{3} = - \frac{2}{3 V_{tb} V_{ts}^*}  \left(C^{bs}_{u u} + 2 C^{bs}_{dd} \right), &\qquad&
\Delta C_{9} = -\frac{4}{3 V_{tb} V_{ts}^*}  \left(C^{bs}_{u u} - C^{bs}_{d d} \right), \nonumber \\
\Delta C_{5} = - \frac{2}{3 V_{tb} V_{ts}^*}  \left(D^{bs}_{u u} + 2 D^{bs}_{d d} \right), &\qquad&
\Delta C_{7} = -\frac{4}{3 V_{tb} V_{ts}^*}  \left(D^{bs}_{u u} - D^{bs}_{d d} \right), \nonumber \\
\Delta \tilde C_{3} = - \frac{2}{3 V_{tb} V_{ts}^*}  \left(\tilde C^{bs}_{u u} + 2 \tilde C^{bs}_{d d} \right), &\qquad&
\Delta \tilde C_{9} = -\frac{4}{3 V_{tb} V_{ts}^*}  \left(\tilde C^{bs}_{u u} - \tilde C^{bs}_{d d} \right), \nonumber \\
\Delta \tilde C_{5} = - \frac{2}{3 V_{tb} V_{ts}^*}  \left(\tilde D^{bs}_{u u} + 2 \tilde D^{bs}_{d d} \right),  &\qquad&
\Delta \tilde C_{7} = -\frac{4}{3 V_{tb} V_{ts}^*}  \left(\tilde D^{bs}_{u u} - \tilde D^{bs}_{d d} \right).  \label{120}
\end{eqnarray}

We pause here to comment on subtleties in Eq.~(\ref{120}).  Recall that in the limit of small fermion mixing angles, Eq.~(\ref{180}) holds for the down-type quarks.  To obtain the CKM matrix as given in Eq.~(\ref{154}) without requiring fine-tuned cancellations, the mixing angles for the up-type left-chiral quarks should also be small in this limit. Due to the $SU(2)_L$ constraint of Eq.~(\ref{152}), therefore, Eq.~(\ref{180}) can also be applied to the up-type left-chiral quarks.  In this case, it is straightforward to see that
\begin{eqnarray}
B_{uu}^L - B_{dd}^L \approx  \tilde \e_{uu}^L - \tilde \e_{dd}^L \equiv 0,
\end{eqnarray}
and hence
\begin{eqnarray}
 &&\Delta C_{9}\approx 0,  \ \ \ \ \ \ \ \ \ \ \ \ \ \ \ \ \ \Delta \tilde C_{7} \approx 0, \nonumber \\ 
 &&\Delta C_{3} \approx - \frac{2}{ V_{tb} V_{ts}^*}  C^{bs}_{dd}, \ \ \ \ 
 \Delta \tilde C_{5} \approx - \frac{2}{ V_{tb} V_{ts}^*}  \tilde D^{bs}_{d d}.
\end{eqnarray}
Note that a relation similar to Eq.(\ref{152}) does not exist for the right-chiral SM fermions, so 
$\Delta C_{7}$ and $\Delta \tilde C_{9}$ are generically non-trivial. 
%It is easy to check that, using the coefficients of no $Z-Z'$ mixing given by Eqs. (\ref{116}), 
%$\Delta C_{3,5,7,9}^1$ are reduced to the results obtained in~\cite{Barger:2004hn}, 
%where $\Delta \tilde C_{3,5,7,9}^1$ are suppressed. 
In regards to the color-allowed penguin operators,  their Wilson coefficients are corrected by $Z'$ effects only at the loop level where the color-indices are mixed by gluons. Since these effects suffer loop and $Z'$ mass double suppressions, 
we will not  consider them further in this paper.

For the $b\to s \bar ll$ transitions,  the $Z'$ contributions to the effective Hamiltonian are
\begin{eqnarray}
{\cal H}_{\rm eff}^{Z'} (b \to s \bar ll) 
&=&  \frac{2G_F}{\sqrt{2}}  \Big( (\bar sb)_{V-A}(C^{bs}_{ll} (\bar ll)_{V-A} + D^{bs}_{ll} (\bar ll)_{V+A}) \nonumber \\ && +  (\bar sb)_{V+A}  (\tilde D^{bs}_{ll} (\bar ll)_{V-A} + \tilde C^{bs}_{ll} (\bar ll)_{V+A}\Big)+ \mbox{h.c.}  \label{119}
\end{eqnarray}
Comparing Eq.~(\ref{119}) with 
\begin{eqnarray} 
\mathcal{H}_{eff}^{Z'}(b \to s \bar l l) &=&
- \frac{G_F}{\sqrt{2}}  V_{tb}^{} V_{ts}^* \Big( \Delta C_{9V} Q_{9V} + \Delta C_{10A} Q_{10A}  \nonumber \\&& + \Delta \tilde C_{9V} \tilde Q_{9V} + \Delta \tilde C_{10A} \tilde Q_{10A} \Big)  + \mbox{h.c.}, \label{130}
\end{eqnarray}
one can see that the $Z'$ corrections to the Wilson coefficients take the following form:
\begin{eqnarray}
\Delta C_{9V} &=&  - \frac{2}{V_{tb} V_{ts}^*} (C^{bs}_{ll}+D^{bs}_{ll}),  \nonumber \\
\Delta C_{10A} &=&  - \frac{2}{V_{tb} V_{ts}^*} (-C^{bs}_{ll}+D^{bs}_{ll}), \nonumber \\ 
\Delta \tilde C_{9V} &=&  - \frac{2}{V_{tb} V_{ts}^*} (\tilde C^{bs}_{ll}+\tilde D^{bs}_{ll}),  \nonumber \\
\Delta \tilde C_{10A} &=&  - \frac{2}{V_{tb} V_{ts}^*} (\tilde C^{bs}_{ll}-\tilde D^{bs}_{ll}).\label{131}
\end{eqnarray}
Note that if the leptons in the process are neutrinos, Eqs.~(\ref{131}) reduces to 
\begin{eqnarray}
\Delta C_{9V} &=&  - \frac{2}{V_{tb} V_{ts}^*} C^{bs}_{ll},  \nonumber \\
\Delta C_{10A} &=&   \frac{2}{V_{tb} V_{ts}^*} C^{bs}_{ll}, \nonumber \\ 
\Delta \tilde C_{9V} &=&  - \frac{2}{V_{tb} V_{ts}^*} \tilde D^{bs}_{ll},  \nonumber \\
\Delta \tilde C_{10A} &=&   \frac{2}{V_{tb} V_{ts}^*} \tilde D^{bs}_{ll}, \label{135}
\end{eqnarray}
since right-handed neutrinos are generally decoupled at low energy scales. 

For $B_s -\bar B_s$ mixing, the $Z'$ corrections to the effective Hamiltonian take the form
\begin{eqnarray}
{\cal H}_{\rm eff}^{Z'}(B_s -\bar B_s)
&=&  \frac{G_F}{\sqrt{2}}  \Big(C^{bs}_{bs} (\bar sb)_{V-A}(\bar sb)_{V-A} +D^{bs}_{bs} (\bar sb)_{V-A}(\bar sb)_{V+A} \nonumber \\&& + \tilde D^{bs}_{bs} (\bar sb)_{V+A}(\bar sb)_{V-A} +\tilde C^{bs}_{bs} (\bar sb)_{V+A}(\bar sb)_{V+A}\Big)+ \mbox{h.c.}  \label{132}
\end{eqnarray}
Once again, upon comparing this expression to 
\begin{eqnarray} 
\mathcal{H}_{eff}^{Z'}(B_s -\bar B_s) =
- \frac{G_F}{\sqrt{2}}  \Big(\Delta C^{ B_s}_1Q^{B_s}_1 +\Delta \tilde C^{ B_s}_1 \tilde Q^{B_s}_1+2\Delta \tilde C^{ B_s}_3 \tilde Q^{B_s}_3  \Big)+ \mbox{h.c.},
 \label{133}
\end{eqnarray} 
the $Z'$ corrections to the Wilson coefficients are easily determined to be
\begin{eqnarray}
\Delta C^{B_s}_1& =&  - C^{bs}_{bs}, \nonumber \\
 \Delta \tilde C^{B_s}_1& =&  - \tilde C^{bs}_{bs} \nonumber \\
  \Delta \tilde C^{B_s}_3& =&  -\frac{1}{2} ( D^{bs}_{bs} + \tilde D^{bs}_{bs} ) =  -  \tilde D^{bs}_{bs}.
 \label{134} 
\end{eqnarray}
As in the $b\to s\bar qq$ transitions, the $Z'$ effects only correct the Wilson coefficients of the color-allowed operators at a higher loop level, so we will not consider them further here. 

To summarize,  in Table (\ref{table1n}) we classify the tree-level $Z'$ contributions to the $b\to s$ transitions according to whether they are relevant or irrelevant to the SM operators. 
\begin{table}[t]
\caption{Classification of the tree-level $Z'$ corrections to the Wilson coefficients in the $b\to s$ transitions.}
\vspace{0.4cm}
\begin{center}
\begin{tabular}{|c|c|c|}
  \hline
 &  SM operators &Beyond SM Operators \\ \hline 
$b \to s \bar qq$ & $\Delta C_a$, $a=3,5,7$ &  $\Delta \tilde C_a$, $a=3,5,9$  \\ \hline
$b \to s \bar ll$ & $\Delta C_a$, $a=9V,10A$& $\Delta \tilde C_a$, $a=9V,10A$
 \\ \hline
 $B_s-\bar B_s$ mixing & $\Delta C^{B_s}_1$ &   $\Delta \tilde C^{B_s}_a$, $a=1,3$  \\ \hline
\end{tabular}
\end{center}\label{table1n}
\end{table}

Before considering the general parameter space, it is worthwhile to consider a few special limits: (1) the LR limit: $|B_{bs}^L| = |B_{bs}^R|$, $\phi_{bs}^L = \phi_{bs}^R$; (2) the LL limit: $\epsilon^{\psi_R} \propto I$; and (3) the RR limit: $\epsilon^{\psi_L} \propto I$, where $I$ is the identity.  The $Z'$ corrections to the Wilson coefficients in these limits are summarized as follows:\\

\noindent {\bf (1) LR limit: $B_{bs}^L = B_{bs}^R$.}
\begin{eqnarray}
 \Delta C^{B_s}_1& =& \Delta \tilde C^{B_s}_1 = \Delta \tilde C^{B_s}_3 =  - (B_{bs}^L)^2, \nonumber \\
\Delta C_{3} &=&   \Delta \tilde C_{5} = - \frac{2}{ V_{tb} V_{ts}^*}  B_{bs}^L B_{dd}^L,  \nonumber \\
\Delta \tilde C_{3} &=& \Delta C_{5}=- \frac{2}{3 V_{tb} V_{ts}^*}  B_{bs}^L \left(B_{u u}^R + 2 B_{dd}^R\right) , \nonumber \\
\Delta C_{7} &=& \Delta \tilde C_{9}  = -\frac{4}{3 V_{tb} V_{ts}^*}  B_{bs}^L\left(B_{u u}^R- B_{d d}^R \right), \nonumber \\
\Delta C_{9V} &=&  \Delta \tilde C_{9V} =- \frac{2}{V_{tb} V_{ts}^*} B_{bs}^L(B_{ll}^L+B_{ll}^R),  \nonumber \\
\Delta C_{10A} &=& \Delta \tilde C_{10A}  =- \frac{2}{V_{tb} V_{ts}^*} B_{bs}^L(-B_{ll}^L+B_{ll}^R). 
 \label{190} 
\end{eqnarray}

\noindent {\bf (2) LL limit: $\epsilon^{\psi_R} \propto I$.}
  \begin{eqnarray}
 \Delta C^{B_s}_1& =& - (B_{bs}^L)^2, \nonumber \\
\Delta C_{3} &=& - \frac{2}{ V_{tb} V_{ts}^*}  B_{bs}^L  B_{dd}^L, \nonumber \\
\Delta C_{5} &=&  - \frac{2}{3 V_{tb} V_{ts}^*}  B_{bs}^L\left(B_{u u}^R + 2 B_{dd}^R \right), \nonumber \\
\Delta C_{7} &=& -\frac{4}{3 V_{tb} V_{ts}^*}  B_{bs}^L\left(B_{u u}^R - B_{d d}^R \right), \nonumber \\
\Delta C_{9V} &=&  - \frac{2}{V_{tb} V_{ts}^*} B_{bs}^L(B_{ll}^L+B_{ll}^R),  \nonumber \\
\Delta C_{10A} &=& - \frac{2}{V_{tb} V_{ts}^*} B_{bs}^L(-B_{ll}^L+B_{ll}^R). 
 \label{191} 
\end{eqnarray}

\noindent {\bf (3) RR limit: $\epsilon^{\psi_L} \propto I$.}
  \begin{eqnarray}
 \Delta \tilde C^{B_s}_1& =& - (B_{bs}^R)^2, \nonumber \\
\Delta \tilde C_{3} &=& - \frac{2}{3 V_{tb} V_{ts}^*}  B_{bs}^R\left(B_{u u}^R + 2 B_{dd}^R \right), \nonumber \\
\Delta \tilde C_{5} &=&  - \frac{2}{V_{tb} V_{ts}^*}  B_{bs}^R B_{dd}^L, \nonumber \\
\Delta \tilde C_{9} &=& -\frac{4}{3 V_{tb} V_{ts}^*}  B_{bs}^R\left(B_{u u}^R - B_{d d}^R \right), \nonumber \\
\Delta \tilde C_{9V} &=& - \frac{2}{V_{tb} V_{ts}^*} B_{bs}^R(B_{ll}^L+B_{ll}^R),  \nonumber \\
\Delta \tilde C_{10A} &=& - \frac{2}{V_{tb} V_{ts}^*} B_{bs}^R(-B_{ll}^L+B_{ll}^R). 
 \label{192} 
\end{eqnarray}    

We will focus on the correlations between $B_s-\bar B_s$ mixing and the hadronic $B_d$ meson decays. For the latter, though $Z^\prime$-mediated effects can occur in both the QCD and EW penguins, we make a conservative assumption in this paper that they are mainly manifest in the EW penguins, such that $|\Delta C_{3,5}|\ll |\Delta C_7|$, as suggested in~\cite{Buras:2003dj,Barger:2003hg}. 
With this restriction, there are only three relevant parameters for each special limit: the modulus of $B_{bs}^L$ (or $B_{bs}^R$), its phase $\phi_{bs}^L$ (or $\phi_{bs}^R$), and the real $B_{dd}^R(\simeq -B_{uu}^R/2)$. These parameters need to satisfy 
\begin{eqnarray}
|B_{bs}^L| < |B_{dd}^L| \ll |B_{dd}^R|
\end{eqnarray}
in the LR and LL limits, and 
\begin{eqnarray}
|B_{bs}^R| < |B_{dd}^R|, \ \ \ \  |B_{dd}^L| \ll |B_{dd}^R|
\end{eqnarray}
in the RR limit. Here $|B_{bs}^{L,R}| < |B_{dd}^{L,R}|$ is due to the fact that, under the assumption of small fermion mixing angles, the modulous of off-diagonal elements in the coupling matrix in Eq. (\ref{160}) should be smaller than that of diagonal ones; $|B_{dd}^L| \ll |B_{dd}^R|$ is due to $|\Delta C_{3,5}|\ll |\Delta C_7|$.
Later in the paper, we will consider the more general parameter space for the $Z'$-mediated effects in the EW penguins, which has five free parameters: $|B_{bs}^{L,R}|$, $\phi_{bs}^{L,R}$ and $B_{dd}^R$.

%%%%%%%%%%%%%%%%%%%%%%%%%%%%%%%%%%%%%%%%%%
\subsection{Effective Couplings at the $b$ Mass Scale}
%%%%%%%%%%%%%%%%%%%%%%%%%%%%%%%%%%%%%%%%%%
To achieve sufficient precision for these observables, it is necessary to have an accurate knowledge of the relevant Wilson coefficients 
at  the $b$ quark mass scale $m_b=4.2$ GeV. The Wilson coefficients at the $b$ mass scale can be obtained as follows:
\begin{equation} 
{\vec{C}} (m_b) = U(m_{b},M_{W})~ {\vec C} (M_{W}),   \label{401}
\end{equation}
where $\vec C$ is a vector with entries consisting of the Wilson coefficients and $U$ is the evolution matrix. The observables can then be expressed in terms of the Wilson coefficients at the $m_b$ scale (for general discussions, see e.g.~\cite{Buchalla:1995vs}).
All parameter values used in our calculations are summarized in~\ref{Parameters}.\\

\noindent $\bullet$ {\bf $B_s$ mixing.}
Following~\cite{Buchalla:1995vs}, the NP probes $C_{B_s}$ and $\phi_{B_s}$, which are defined in Eq.~(\ref{001}),
are calculated to be 
\begin{eqnarray}
C_{B_s} e^{2i\phi_{B_s}} &=& 1 - 3.59 \times 10^5  (\Delta C_1^{B_s} + \Delta \tilde C_1^{B_s}) 
+ 2.04 \times 10^6 \Delta \tilde C_3^{B_s} \label{403}
\end{eqnarray}
at the $m_b$ scale. The large coefficients of the correction terms are due to the fact that the NP is introduced at tree-level while the SM limit is a loop-level effect.\\

\noindent $\bullet$  {\bf $B_d \to \pi K_S$ decays.}
%It is known that non-penguin dominated effects do provide some corrections to ${\mathcal S}_{\pi KS}$~\cite{Chiang:2004nm,Beneke:2005pu,Cheng:2005bg} which tend to increase its value 
%from $\sin2\beta$ at the pure penguin limit to around $0.8$. Recently, using isospin relations, it is predicted~\cite{Fleischer:08wb,Gronau:08gu} that an larger value 
%\begin{eqnarray}
%{\mathcal S}_{\pi KS} = 0.99^{+0.01}_{-0.261}  \label{404}
%\end{eqnarray}
%is favored in the SM, which is about two standard deviations away from the experimental value in Table {\ref{table2}}.  
The $B_d \to \pi K_S$ decays have recently received considerable interest in the literature (see e.g.~\cite{Buras:2003dj,Fleischer:2008wb,Gronau:2008gu,Chiang:2004nm,Baek:2009pa,Ciuchini:2008eh}).  In~\cite{Buras:2003dj}, it is pointed out that a deviation of ${\cal S}_{\pi K_S}$ from its SM value can be understood as a modification of the ratio
\begin{eqnarray}
q e^{i\phi} = \frac {P} {T+C}, \label{405}
\end{eqnarray}
in which $T$, $C$ and $P$ denote the color-allowed tree, color-suppressed tree, and EW penguin contributions in the decay amplitude, respectively. 
In the class of models considered here, the family non-universal $Z'$ interactions modify $qe^{i\phi}$ through the relation
\begin{eqnarray}
q e^{i\phi} &=& 0.76 (1 + 158.1 \Delta C_7 -102.4 \Delta \tilde C_9),  \label{406}
\end{eqnarray}
in which $q$ and $\phi$ are given by 0.76 and zero, respectively, in the SM limit. \\

\noindent $\bullet$ {\bf $B_d \to (\psi, \phi, \eta', \rho, \omega, f^0)K_S$ decays.}
The direct and the mixing-induced $CP$ asymmetries in the $B_d$ hadronic decays are parametrized as follows: 
\begin{eqnarray}
{\mathcal C}_{f_{CP}}
= \frac{1-|\lambda_{f_{CP}}|^2}{1+|\lambda_{f_{CP}}|^2} ~,
\qquad
{\mathcal S}_{f_{CP}}
= \frac{2{\rm Im} \left[ \lambda_{f_{CP}} \right]}{1+|\lambda_{f_{CP}}|^2}.
\end{eqnarray}
In the above, $\lambda_{f_{CP}}$  is defined by
\begin{eqnarray}
\lambda_{f_{CP}}
\equiv -\eta_{f_{CP}} \frac{q_{B_d}}{p_{B_d}}
       \frac{\bar{A}_{f_{CP}}}{A_{f_{CP}}},  \label{lambda}
\end{eqnarray}
with 
\begin{eqnarray}
\frac{q_{B_d}}{p_{B_d}} \Big|_{\Delta \Gamma_{B_d}=0}=-\frac{(M_{B_d})_{12}^*}{|(M_{B_d})_{12}|}= -e^{-2i\phi_{B_d}}.
\end{eqnarray}
Here $q_{B_d}$ and $p_{B_d}$ are $B_d$ mixing coefficients 
\begin{eqnarray}
| B_L \rangle &=& p_{B_d} | B_d \rangle + q_{B_d} | \bar B_d \rangle \nonumber \\ 
| B_H \rangle &=& p_{B_d} | B_d \rangle - q_{B_d} | \bar B_d \rangle,
\end{eqnarray}
$M_{B_d}$ is the $B_d-\bar B_d$ mass matrix, and $A_{f_{CP}}$ is the decay amplitude of $B_d \to f_{CP}$ ($\bar A_{f_{CP}}$ is its $CP$ conjugate.).

The SM predicts that $\phi_{B_d} = \beta \equiv \arg\left[-(V_{cd}V_{cb}^*)/(V_{td}V_{tb}^*)\right]$ and that a non-trivial weak phase enters $A_{f_{CP}} $ only at order ${\mathcal O}(\lambda^2)$.   Therefore, for time-dependent decays proceeding via $b\to s \bar qq(q=u,d,c,s)$, including $B_d\to \psi K_S$ and penguin-dominated modes such as $B_d \to (\phi, \eta', \pi, \rho, $ $\omega, f^0 ) K_S$, the relations in Eq. (\ref{TDCP}) are obtained.
%\begin{eqnarray}
%-\eta_{f_{CP}}{\mathcal S}_{f_{CP}} = \sin 2 \beta + {\mathcal O}(\lambda^2),  \ \ \ \  {\mathcal C}_{f_{CP}} = 0 + {\mathcal O}(\lambda^2). 
%\end{eqnarray}
%where $\eta_{f_{CP}}=\pm1$ is the CP eigenvalue for the final state $f_{CP}$.
However, these results are greatly changed with the involvement of family non-universal $Z'$ bosons, since this allows for a new weak phase to enter $A_{f_{CP}}$ at tree level. Following Ali~et.~al.~\cite{Ali:1998eb},  the 
$\lambda_{f_{CP}}$ parameters of $B_d \to (\psi, \phi, \eta', \pi, \rho, \omega, f^0)K_S$ are given by
\begin{eqnarray} 
\lambda_{\psi K_S} &=&(-0.63+ 0.74 i)   \label{407}\\ \nn && \frac{1 +
  (0.18 - 0.01 i )(\Delta C_7 + \Delta \tilde C_7)^{*} -
  (0.06 - 0.04 
  i)(\Delta C_9 + \Delta \tilde C_9)^{*}}{ 1 + (0.17 + 0.01 i
  )(\Delta C_7 + \Delta \tilde C_7) - (0.05 + 0.05 i)(\Delta C_9 +
  \Delta \tilde C_9)}~,  
\end{eqnarray} 
\begin{eqnarray} 
\lambda_{\phi K_S} &=& (-0.70 + 0.70 i)  \label{408}\\ \nn && \frac{1 +
 (14.57 + 5.88 i) (\Delta C_7 + \Delta \tilde C_7)^{*} + (15.08 +
 5.92 i) (\Delta C_9 + \Delta \tilde C_9)^{*}}{ 1 + (14.39 + 5.64 i)
 (\Delta C_7 + \Delta \tilde 
        C_7) + (14.90 + 5.66 i)
        (\Delta C_9 + \Delta \tilde C_9) }~, 
\end{eqnarray} 
\begin{eqnarray} 
\lambda_{\eta^\prime K_S} &=& (-0.70 + 0.69 i)   \label{409}\\ \nn 
&& \frac{1 +
  (2.11 + 0.67 i)(\Delta C_7 + \Delta \tilde C_7)^{*} + (2.10 +
  0.54 i)(\Delta C_9 + \Delta \tilde C_9)^{*}}{1 + (2.08 + 0.65
  i)(\Delta C_7 + \Delta \tilde C_7) + (2.07 + 0.52 i)(\Delta C_9 +
  \Delta \tilde C_9)}~, 
\end{eqnarray}
%\begin{eqnarray} 
%\lambda_{\eta K_S} &=& (-0.67 + 0.74 i)  \label{410}\\ \nn 
%&& \frac{1 - (30.78 + 11.37 i)(\Delta C_7 + \Delta \tilde C_7)^{*} + (35.23 + 12.89
 % i)(\Delta C_9 + \Delta \tilde C_9)^{*}}{1 - (31.50 + 9.54
 % i)(\Delta C_7 + \Delta \tilde C_7) + (36.05 + 10.79 i)(\Delta C_9 +
 % \Delta \tilde C_9)}~, 
%\end{eqnarray}
\begin{eqnarray} 
\lambda_{\rho K_S} &=& (-0.74 + 0.66 i)  \label{411}\\ \nn && \frac{1 -
  (38.75 + 3.29 i )(\Delta C_7 + \Delta \tilde C_7)^{*} - (47.95 + 4.11
  i)(\Delta C_9 + \Delta \tilde C_9)^{*}}{1 - (38.11 + 5.23 i
  )(\Delta C_7 + \Delta \tilde C_7) - ( 47.15 + 6.50 i)(\Delta C_9 +
  \Delta \tilde C_9)}~, 
  \end{eqnarray} 
\begin{eqnarray} 
\lambda_{\omega K_S} &=&  (-0.71 + 0.70 i)  \label{412}\\ \nn && \frac{1 +
  (31.97 + 4.76 i )(\Delta C_7 + \Delta \tilde C_7)^{*} + (18.84 + 2.75
  i)(\Delta C_9 + \Delta \tilde C_9)^{*}}{ 1 + (31.81 + 4.67 i
  )(\Delta C_7 + \Delta \tilde C_7) + (18.74 + 2.70 i)(\Delta C_9 +
  \Delta \tilde C_9)}, 
\end{eqnarray} 
\begin{eqnarray} 
\lambda_{f^0 K_S} &=&  (-0.70 + 0.70 i)  \label{413}\\ \nn && \frac{1 +
  (3.19 + 0.93 i )(\Delta C_7 + \Delta \tilde C_7)^{*} - (0.12 + 0.15
  i)(\Delta C_9 + \Delta \tilde C_9)^{*}}{1 + (3.16 + 0.90 i
  )(\Delta C_7 + \Delta \tilde C_7) - ( 0.12 + 0.15 i)(\Delta C_9 +
  \Delta \tilde C_9)}~. 
\end{eqnarray} 
In contrast to the $B_d\to \psi K_S$ decay, in which the NP
effects are suppressed by the SM tree-level contribution, family non-universal $U(1)^\prime$ couplings indeed result in sizable corrections to $\lambda_{f_{CP}}$ for the penguin-dominated modes.

%%%%%%%%%%%%%%%%%%%%%%%%%%%%%%%%%%%%%%%%%%%%%
\section{Results and Analysis}
\subsection{Correlated Analysis (I) -- Special Limits}
%%%%%%%%%%%%%%%%%%%%%%%%%%%%%%%%%%%%%%%%%%%%%
\label{SL}

\begin{figure}[ht]
\begin{center}
\includegraphics[width=0.45\textwidth]{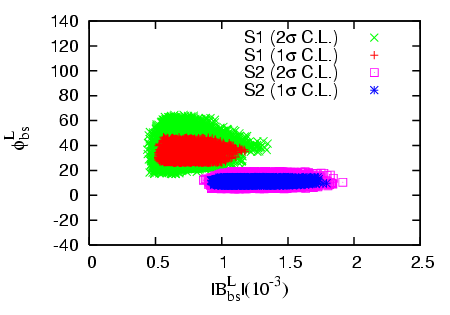}
\includegraphics[width=0.45\textwidth]{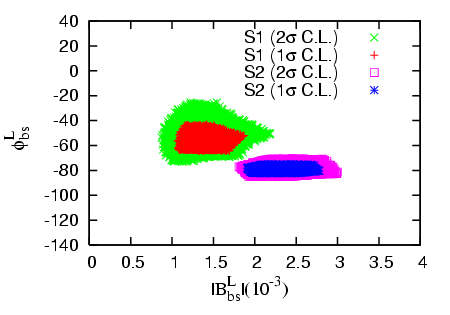}
\caption{Correlated constraints on $|B_{bs}^L|$ and $\phi_{bs}^L$ are presented.  In these two panels, random values  for $C_{B_s}$ and $\phi_{B_s}^{\rm NP}$ from the experimentally allowed regions (see Table~\ref{table1}) are mapped to the $|B_{bs}^L|-\phi_{bs}^L$ plane using Eq.~(\ref{451}), with an assumed $25\%$ uncertainty (a typical value from non-perturbative effects) assumed for the coefficients. The left (right) panel is the LR (LL) limit.} %LABEL SIZES NEED TO BE ENLARGED.
\label{SLfigure1}
\end{center}
\end{figure}

\begin{figure}[ht]
\begin{center}
\includegraphics[width=0.45\textwidth]{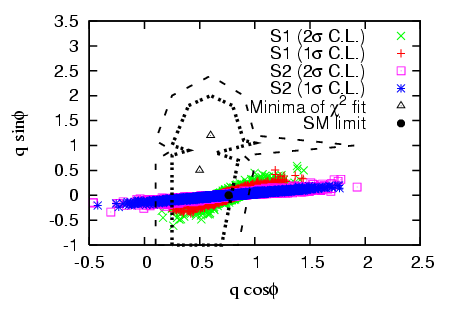}
\includegraphics[width=0.45\textwidth]{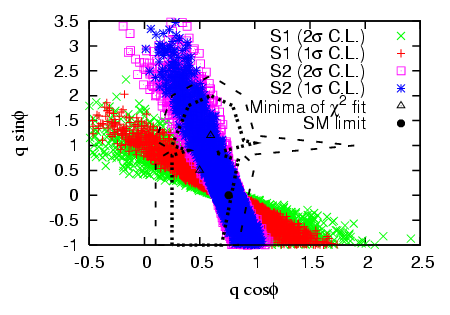}
\caption{The constraints on $B_{dd}^R$ from $q e^{i\phi}$ are shown. The points from the $|B_{bs}^L| - \phi_{bs}^L$ plane (see Fig.~\ref{SLfigure1}) are randomly combined with 
scattered points of $B_{dd}^R$ ($10^{-3} \le |B_{dd}^R| \le 10^{-1}$) and then mapped to the $q\cos\phi -q\sin\phi$ plane according to Eq.~(\ref{452}). The colors of the points in this plane indicate the C.L. that their inverse images represent in Fig.~\ref{SLfigure1}.  The two dashed lines specify the experimentally allowed ranges that result from the $\chi^2$ fit of the $B\to\pi K$ (and $B\to\pi\pi$) data at $1 \sigma$ and $90\%(\simeq 1.7\sigma)$ C.L., respectively~\cite{Fleischer:2008wb}. The left (right) panels are the LR (LL) limits.}
\label{SLfigure2}
\end{center}
\end{figure}

\begin{figure}[ht]
\begin{center}
\includegraphics[width=0.45\textwidth]{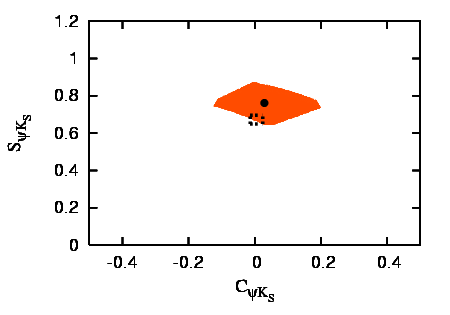}
\includegraphics[width=0.45\textwidth]{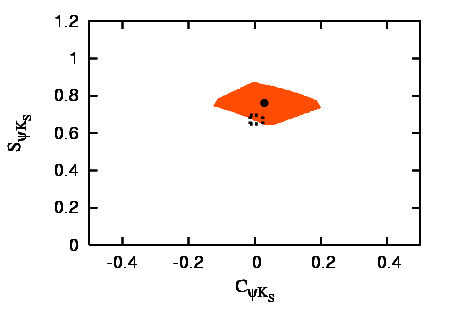}
\caption{The time-dependent $CP$ asymmetries of the charmed $B_d\to \psi K_S$ decay are presented (with $|V_{ub}|=3.51\times 10^{-3}$, as is used in the SM calculation~\cite{CKMfitter08}). The box is at $1\sigma$ C.L. and the dark point is the SM limit.  The left (right) panels are the LR (LL) limits.}
\label{SLfigure4}
\end{center}
\end{figure}

\begin{figure}[ht]
\begin{center}
\includegraphics[width=0.35\textwidth]{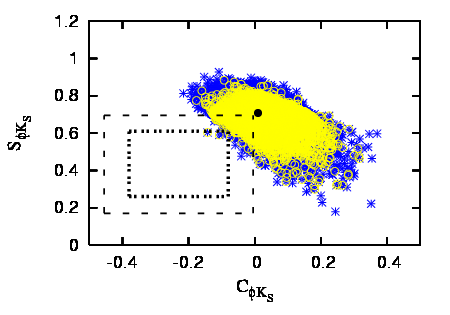}
\includegraphics[width=0.35\textwidth]{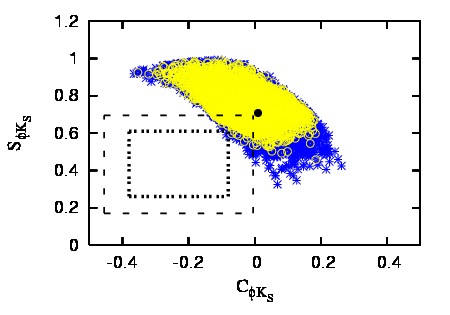}
\includegraphics[width=0.35\textwidth]{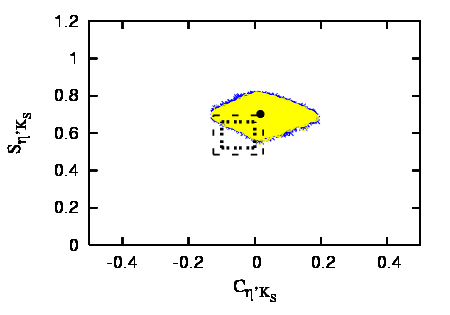}
\includegraphics[width=0.35\textwidth]{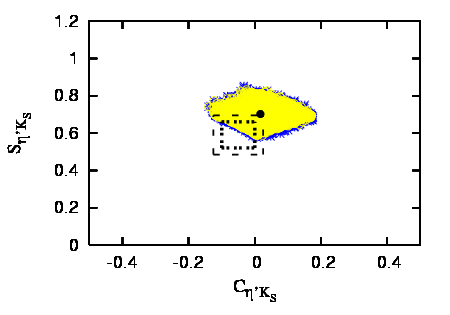}
\includegraphics[width=0.35\textwidth]{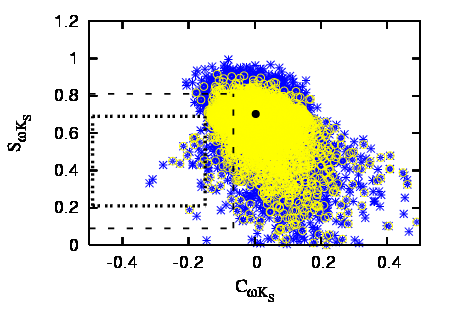}
\includegraphics[width=0.35\textwidth]{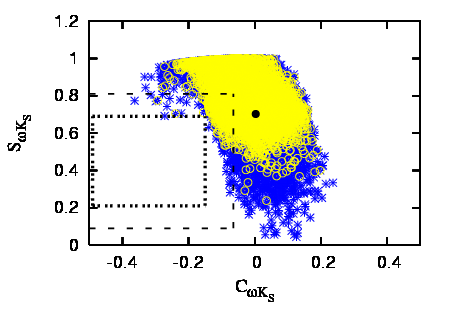}
\includegraphics[width=0.35\textwidth]{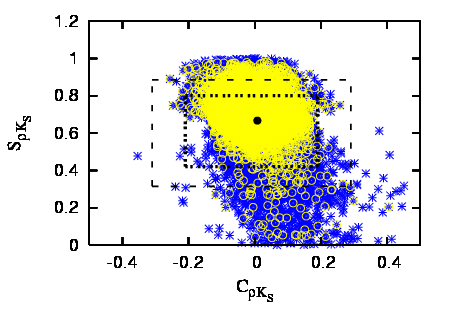}
\includegraphics[width=0.35\textwidth]{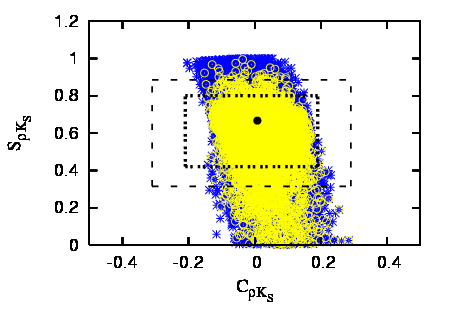}
\includegraphics[width=0.35\textwidth]{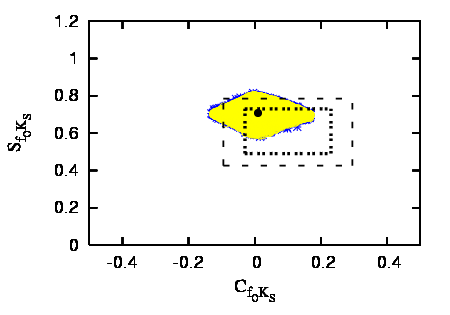}
\includegraphics[width=0.35\textwidth]{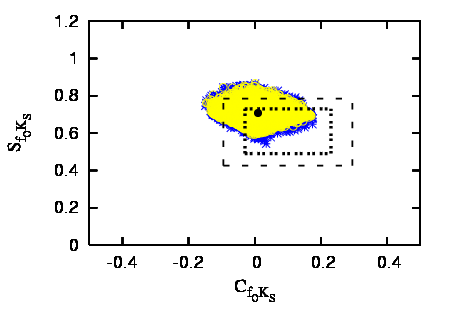}
\caption{With $B_{bs}^L$ and $B_{dd}^R$ constrained by $B_s-\bar B_s$ mixing and $B_d\to \pi K_S$, the NP contributions to  ${\mathcal C}_{(\phi, \eta', \rho, \omega, f_0)K_S}$ and ${\mathcal S}_{(\phi, \eta', \rho, \omega, f_0)K_S}$ are shown. The left (right) panels are the LR (LL) limits.  The colors specify the C.L. that their inverse image points represent in Figs.~\ref{SLfigure1} and~\ref{SLfigure2} (yellow for $1 \sigma$ C.L. and blue for $2 \sigma$ and $1.7 \sigma$ C.L.).  The boxes specify the experimentally allowed regions at 1$\sigma$ and $1.7\sigma$, and the dark point denotes the SM limit.}
\label{SLfigure3}
\end{center}
\end{figure}

\begin{figure}[ht]
\begin{center}
\includegraphics[width=0.45\textwidth]{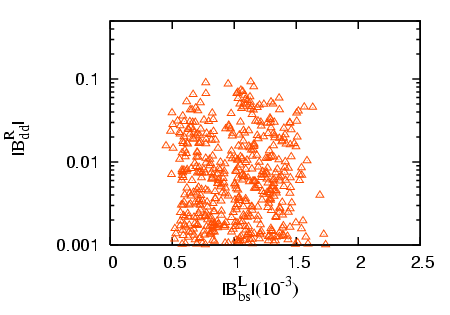}
\includegraphics[width=0.45\textwidth]{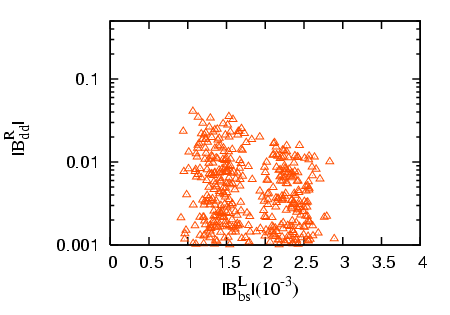}
\includegraphics[width=0.45\textwidth]{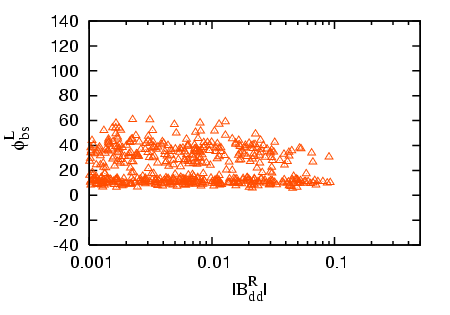}
\includegraphics[width=0.45\textwidth]{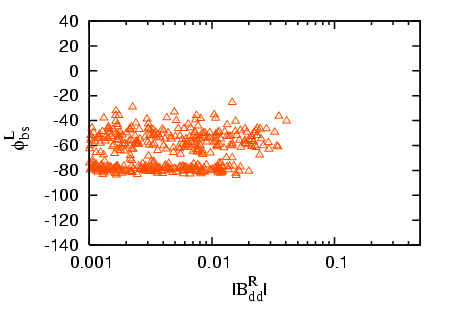}
\includegraphics[width=0.45\textwidth]{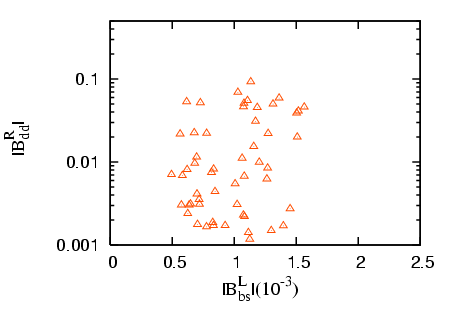}
\includegraphics[width=0.45\textwidth]{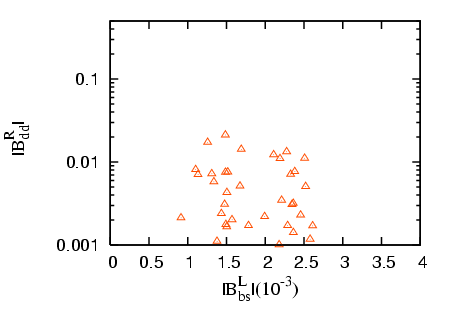}
\includegraphics[width=0.45\textwidth]{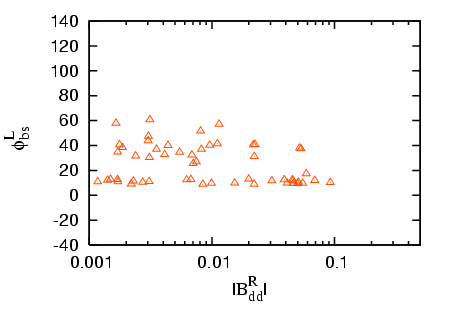}
\includegraphics[width=0.45\textwidth]{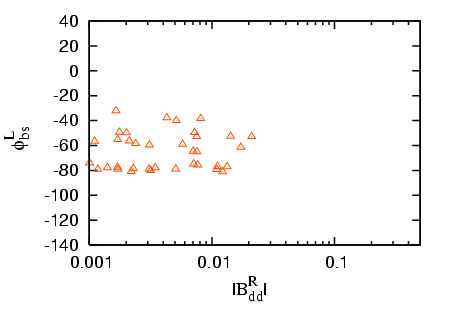}
\caption{The allowed $|B_{bs}^L|$, $\phi_{bs}^L[^\circ]$ and $B_{dd}^R$ are shown. They are constrained from $B_s-\bar B_s$ mixing (at $2 \sigma$ C.L.) and the $\chi^2$ fit of the $B\to\pi K$ (and $B\to\pi\pi$) data (at $1 \sigma$ C.L.), then selected by  ${\mathcal C}_{(\phi, \eta', \rho, \omega, f_0)K_S}$, ${\mathcal S}_{(\phi, \eta', \rho, \omega, f_0)K_S}$ (at $1.7 \sigma$ C.L. for the first four panels and $1.5\sigma$ C.L. for the others). The left (right) panels are the LR (LL) limits.}  %The uncertainties due to non-perturbative physics in the SM and NP calculations are assumed to be $15\%$ and $25\%$, respectively.}
\label{SLfigure5}
\end{center}
\end{figure}

\begin{figure}[ht]
\begin{center}
\includegraphics[width=0.45\textwidth]{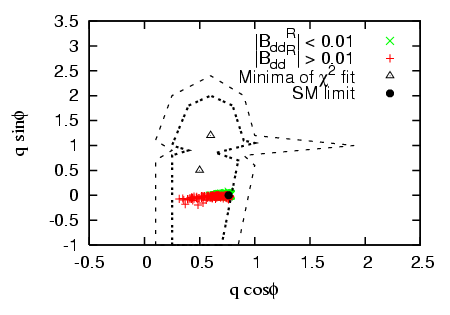}
\includegraphics[width=0.45\textwidth]{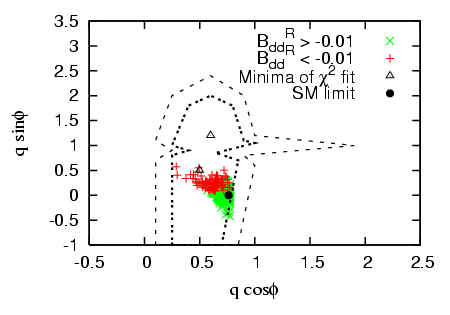}
\includegraphics[width=0.45\textwidth]{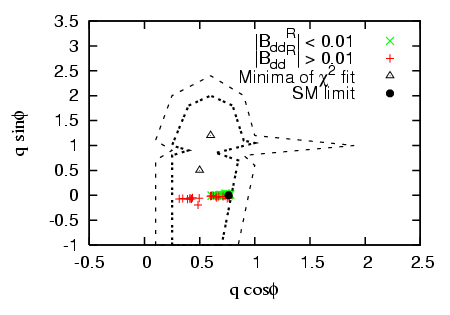}
\includegraphics[width=0.45\textwidth]{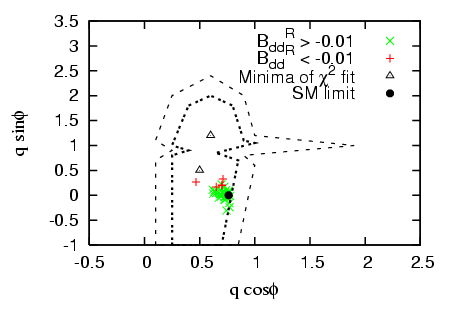}
\caption{The points in Fig.~\ref{SLfigure5} are inversely mapped to the $q\cos\phi-q\sin\phi$ plane. The panels in the first row correspond to a parameter selection from ${\mathcal C}_{(\phi, \eta', \rho, \omega, f_0)K_S}$, ${\mathcal S}_{(\phi, \eta', \rho, \omega, f_0)K_S}$ at $1.7 \sigma$ C.L. in Fig.~\ref{SLfigure5} , and those in the second row at $1.5\sigma$ C.L.. The left (right) panels are the LR (LL) limits. }
\label{SLfigure6}
\end{center}
\end{figure}

In this section, we will present a correlated analysis of the $\Delta B =1, 2$ processes which occur via $b\to s$ transitions, focusing first on the special limits of the parameter space as presented in Section~\ref{section 2}.  As the physics of the RR limit is very similar to that of the LL limit, we focus in this paper on the LR and LL limits as representative examples.  

We first consider the constraints on this class of  family non-universal $U(1)^\prime$ scenarios which arise from $B_s-\bar{B}_s$ mixing. With the renormalization scale chosen as the $b$-quark mass, $m_b=4.2$ GeV, the NP probes  $C_{B_s}$ and $\phi_{B_s}^{\rm NP}$ are given by
\begin{eqnarray}
C_{B_s} e^{2i\phi_{B_s}^{\rm NP}} &=& 1 + 1.32 \times 10^6  \Delta C_1^{B_s}  \nonumber \\
C_{B_s} e^{2i\phi_{B_s}^{\rm NP}} &=& 1 - 3.59 \times 10^5  \Delta \tilde C_1^{B_s}  \label{451}
\end{eqnarray}
in the LR and LL limits, respectively.  These conditions involve two of the three free parameters of each limit: $|B_{bs}^L|$ and $\phi_{bs}^L$. The experimental constraints on these parameters from $B_s-\bar{B}_s$ mixing are illustrated in  Fig.~\ref{SLfigure1}. The left panel corresponds to the LR limit and the right one corresponds to the LL limit; in this section we will present the results for these two limits together, so that it is easy to make comparisons between the two cases. In each case, there are two separate shaded regions, corresponding to the two $\phi_{B_s}^{\rm NP}$ solutions (see Table~\ref{table1}). For each region,  the various colors of the points specify the different confidence levels (C.L.) of the relevant $C_{B_s}$ and $\phi_{B_s}^{\rm NP}$ values. To explain the observed discrepancy in $B_s-\bar B_s$ mixing from the SM prediction, $|B_{bs}^L|$ is required to be $\sim 10^{-3}$.  This reflects two facts: (1) unlike $\phi_{B_s}^{\rm NP}$, the modulus $C_{B_s}$ does not deviate from its SM prediction significantly (the experimental value of $C_{B_s}$ has an at most ${\mathcal O}(1)$ shift from its SM prediction at $2\sigma$ C.L.); (2) the $Z'$ corrections are from tree level, and hence can easily explain this small deviation (only a small coupling is necessary, according to Eq. (\ref{451}) and Eq. (\ref{190}), Eq. (\ref{191})). 
The smallness of $|B_{bs}^L|$ is generically consistent with our assumption of  small fermion mixing angles, since $B_{bs}^L$ is proportional to them as well as to $g_2 M_Z/(g_1 M_{Z'})$ (see Eq. (\ref{180}) and the comments under Eq. (\ref{116})). In addition, due to the smallness of $|B_{bs}^L|$, the experimental constraints from the branching ratio ${\rm Br}(B_s \to \mu^+ \mu^-)$ can be easily satisfied. For details, see  \ref{Bstomumu}. 

Before move to the other $b\to s$ processes, we have some comments on the influence of $B_s-\bar B_s$ mixing on another FCNC $Z'$ coupling $B^{L,R}_{bd}$. In the SM, the mass differences of $B_d$ and $B_s$ masons are predicted to be (e.g., see~\cite{Nierste:2006na})
\begin{eqnarray}
\Delta M_d^{\rm SM} &=& (0.53\pm 0.02)\left(\frac{|V_{td}|}{0.0082} \right)^2 \left(\frac{f_{B_d}}{200 {\rm MeV}}\right)^2 \frac{B}{0.85}  {\rm ps}^{-1} \nonumber \\
\Delta M_s^{\rm SM} &=& (19.3\pm 0.6) \left(\frac{|V_{ts}|}{0.00405} \right)^2 \left(\frac{f_{B_s}}{240 {\rm MeV}}\right)^2 \frac{B}{0.85} {\rm ps}^{-1} 
\end{eqnarray}
with $f_{B_d, B_s}$ being decay constant of  $B_d (B_s)$ and $B$ being bag factor. Comparing with the experimental data~\cite{Barberio:2007cr} 
\begin{eqnarray}
\Delta M_d &=& \Delta M_d^{\rm SM} (1+\frac{\Delta M_d^{\rm NP}}{\Delta M_d^{\rm SM}}) = 0.507 \pm 0.005 {\rm ps}^{-1}  \nonumber \\
\Delta M_s &=& \Delta M_s^{\rm SM} (1+\frac{\Delta M_s^{\rm NP}}{\Delta M_s^{\rm SM}})  =  17.77 \pm 0.12 {\rm ps}^{-1},
\end{eqnarray}
we have the relation 
\begin{eqnarray}
\frac{\Delta M_{d}^{\rm NP}}{\Delta M_{d}^{\rm SM}} \sim \frac{\Delta M_{s}^{\rm NP}}{\Delta M_{s}^{\rm SM}}
 \end{eqnarray}
with
 \begin{eqnarray}
\frac{\Delta M_{d}^{\rm NP}}{\Delta M_{s}^{\rm NP}} \sim \frac{|B_{bd}^{L,R}|^2}{|B_{bs}^{L,R}|^2}, \ \ \ \  \frac{\Delta M_{d}^{\rm SM}}{\Delta M_{s}^{\rm SM}} \sim \lambda^2 \approx 0.04.
\end{eqnarray}
But, according to Eq. (\ref{180}), the modulous of $B^{L,R}_{bd}$ usually is comparable with that of $B^{L,R}_{bs}$,
a fine-tuning of ${\mathcal O} (10\%)$ level therefore is needed in $|B^{L,R}_{bd}|$ to satisfy the experimental constraints from $\Delta M_d$. In this paper, we will work under the assumption of negligible $|B^{L,R}_{bd}|$, and therefore will neglect its possible effect in the NP observables.

The second process of interest is $B_d \to \pi K_S$. The time-dependent $CP$ asymmetries of these decays can be sizably affected by NP, as has been pointed out in~\cite{Buras:2003dj}.  The experimental constraints on $q e^{i\phi} $ (defined in Eq. (\ref{405})) for different C.L.'s from the $B\to\pi K$ (and the $B\to\pi\pi$) data have previously been obtained in~\cite{Fleischer:2008wb}.
In Fig.~\ref{SLfigure2},  we illustrate how $B_{dd}^R$ is constrained through $q e^{i\phi}$, using the parameter values of $|B_{bs}^L|$ and $\phi_{bs}^L$ obtained in Fig.~\ref{SLfigure1}, along with the following relations:
\begin{eqnarray}
q e^{i\phi} &=& 0.76 (1 + 55.7 \Delta C_7) \nonumber \\
q e^{i\phi} &=& 0.76 (1+158.1\Delta  C_7), \label{452}
\end{eqnarray}
which are valid in the LR and LL limits, respectively. There are two distribution regions which are specified by different colors in each panel,  again due to the two $\phi_{B_s}^{\rm NP}$ solutions. Note that in the LL limit, the shaded region passes through both of the minimal points that were found in the $\chi^2$ fit of the $B\to\pi K$ and $B\to\pi\pi$ data in~\cite{Fleischer:2008wb}.

In the scenarios under consideration, the constraints from $B_s-\bar B_s$ mixing and $B_d \to \pi K_S$ decays place bounds on each of the three free parameters $|B_{bs}^L|$, $\phi_{bs}^L$ and $B_{dd}^R$. The natural question is then whether the experimentally allowed values for these parameters also satisfy the constraints resulting from the possibly anomalous values of $\Delta {\mathcal C}_{f_{CP}}$ and $\Delta {\mathcal S}_{f_{CP}}$ in the remaining penguin-dominated $B_d \to (\phi, \eta', \rho, \omega, f_0)K_S$ decays.  To address this issue,  we assume a $15\%$ uncertainty in the SM calculations for each of these modes (as well as for the $B_d\to \pi K_S$ mode) and a $25\%$ uncertainty for the NP contributions. Here $15\%$ is a typical uncertainty level for the hadronic matrix elements of the SM FC operators (e.g., see~\cite{Wirbel:1985ji}) and is also the least necessary one to explain the experimental data of ${\mathcal C}_{\psi K_S}$ and ${\mathcal S}_{\psi K_S}$ in the SM (see Fig.~\ref{SLfigure4} where the NP effects are negligible). As for the difference of the uncertainty levels between the SM and NP calculations, it is caused by the fact that the hadronic matrix elements of the FC operators in the SM are better understood than they are for the NP operators. In Fig.~\ref{SLfigure3}, we systematically illustrate the time-dependent $CP$ asymmetries in the penguin-dominated modes, using the parameter values obtained above and Eqs.~(\ref{407})--(\ref{413}). For these modes with the exception of $B_d \to \rho K_S$, there are $0.5\sim 2 \sigma$ deviations for ${\mathcal C}_{f_{CP}}$, ${\mathcal S}_{f_{CP}}$, or both. Though our model only induces negligible effects on $B_d -\bar B_d$ mixing under the assumption of small $|B_{bd}|$, due to the interference effects between the $B_d-\bar B_d$ mixing phase and $\phi_{bs}^L$ which affects the decay asymmetries $\frac{\bar{A}_{f_{CP}}}{A_{f_{CP}}}$ in Eq. (\ref{lambda}), the points in Fig.~\ref{SLfigure3} are scattered away from the SM limits. This results in a dispersion such that there are always some points lying in the $1\sigma$ region for each of these modes.  

To show that all of the constraints can be satisfied simultaneously, we have carried out a correlated analysis among the $B_s-\bar B_s$ mixing and the $B_d \to (\pi, \phi, \eta', \rho, \omega, f_0)K_S$  $CP$ asymmetries. The allowed values for $|B_{bs}^L|$, $\phi_{bs}^L$ and $B_{dd}^R$ in the LR and LL limits are illustrated in Fig.~\ref{SLfigure5}. We see that in Fig.~\ref{SLfigure5} there indeed exist parameter regions where the anomalies in $B_s - \bar B_s$ mixing and the time-dependent  $CP$ asymmetries of $B_d \to (\pi, \phi, \eta', \rho, \omega, f_0)K_S$, can be explained by NP at reasonable C.L.. The allowed $|B_{bs}^R|$ and $\phi_{bs}^L$ values can explain both solutions of $B_s - \bar B_s$ mixing phase; and the allowed $|B_{dd}^R|$ values vary from $0.08$ to smaller values. If we want to get a better fit for ${\mathcal C}_{(\pi,\phi, \eta', \rho, \omega, f_0)K_S}$ and ${\mathcal S}_{(\pi,\phi, \eta', \rho, \omega, f_0)K_S}$, $|B_{dd}^R|\simgt 10^{-2}$ is typically required. 
%Since $\Delta C_7$ and $\Delta \tilde C_9$ are proportional to $|B_{dd}^R|$, for points with $|B_{dd}^R| \ll 10^{-2}$, the NP effects are negligible in  ${\mathcal C}_{(\pi,\phi, \eta', \rho, \omega, f_0)K_S}$ and ${\mathcal S}_{(\pi,\phi, \eta', \rho, \omega, f_0)K_S}$. However, this does not necessarily imply that NP is not relevant for them. 
To see this point, we take for example ${\mathcal C}_{\pi K_S}$ and ${\mathcal S}_{\pi K_S}$, and map the points in Fig.~\ref{SLfigure5} back to the $q\cos\phi-q\sin\phi$ plane, as illustrated in Fig~\ref{SLfigure6}. In this figure we see that the points with $B_{dd}^R < -0.01$ are closer to the minima of the $\chi^2$ fit of the $B\to \pi K$ and  $B\to \pi \pi$ data, leading to a better fit compared to the one obtained in the SM limit. Therefore, $|B_{dd}^R|\simgt 10^{-2}$ is important in improving the agreement with experimental data in the penguin-dominated $B_d$ decays.\footnote{This effect can also be seen by requiring a smaller C.L. for the fit of ${\mathcal C}_{(\phi, \eta', \rho, \omega, f_0)K_S}$ and ${\mathcal S}_{(\phi, \eta', \rho, \omega, f_0)K_S}$, which has been shown in the LR limit in Fig. 4 of~\cite{Barger:2009eq}.} 

The favored parameter values for $|B_{dd}^R|$ are interesting for collider detection. For $(V_{d_R} \tilde \epsilon^{d_R} V_{d_R})_{11} \sim {\mathcal O}(1)$, this implies that $g_1M_{Z'}/(g_2 M_{Z}) \sim 10-100$ or a TeV scale  $Z'$ boson for $g_2\simlt g_1$, a range approachable at the LHC (e.g., see~\cite{Langacker:2008yv, Carena:2004xs}). This fact is also important for the effective Lagrangian in Eq. (\ref{111}) which is obtained by integrating out the $Z'$ boson. While applying it to our analysis, we neglected the effects of the renormalization group running between $Z'$ mass scale and EW scale, which is justified only for a small gap between these two scale or for a low-scale $Z'$ boson. 
%Actually, for a $Z'$ boson of large mass scale, the coupling structure introduced in Eq. (\ref{160}) also becomes unconstrained. A TeV scale  $Z'$ boson therefore is highly preferred. Indeed, from the results obtained from above we see that a low-scale $Z'$ is generically favored.
In addition, we emphasize that the favored parameter regions are consistent with our assumption that the non-universal $Z'$ effects in QCD penguins are negligible. This assumption requires 
$|\Delta C_{3,5}| \ll |\Delta C_7|$ or 
%\begin{eqnarray}
$|B_{bs}^L| < |B_{dd}^L| \ll |B_{dd}^R|$.
%\end{eqnarray}
Since $|B_{bs}^L|$ and $|B_{dd}^R|$ are favored to be $\sim 10^{-3}$ and $\simgt 10^{-2}$ respectively, this relation can be easily accommodated. At last, to implement our discussions, we take a $\chi^2$ fit in the SM and in the non-universal $U(1)'$ models for all relevant observables except $q e^{i\phi}$.
%$C_{B_s}$, $\phi_{B_s}^{\rm NP}$, ${\mathcal C}_{(\phi, \eta', \rho, \omega, f_0)K_S}$ and ${\mathcal S}_{(\phi, \eta', \rho, \omega, f_0)K_S}$ 
We find that  the reduced $\chi^2$ value ($i.e.$, $\chi^2/{\rm D.O.F.}$) in the SM is larger than 2, and that of the best fit in both LL and LR limits in the $U(1)'$ models is smaller than 1. 
Therefore, a better fit is obtained in the latter.

\subsection{Correlated Analysis (II) -- General Case}
\label{GC}

%\begin{figure}[ht]
%\begin{center}
%\includegraphics[width=0.60\textwidth]{Figures/Bsmixing_GR.png}
%\caption{The constraints arising from $C_{B_s}$ and $\phi_{B_s}^{NP}$ on $|B_{bs}^{L,R}|$ and $\phi_{bs}^{L,R}$ are shown. 
%in $B_s-\bar B_s$ mixing. 
%The values for $|B_{bs}^{L,R}| (10^{-4}<|B_{bs}^{L,R}|<10^{-2})$  and $\phi_{bs}^{L,R} (-\pi< \phi_{bs}^{L,R} < \pi )$ are randomly generated and then mapped to the $C_{B_s}-\phi_{B_s}^{\rm NP}$ plane using Eq. (\ref{403}). The boxes on the left correspond to the $\phi_{B_s}^{NP}$ solution ``S1'' and the ones on the right correspond to the solution ``S2'' (see Table~\ref{table2}). For each set, the black (inside) box and the grey (outside) box denote $1\sigma$ and $2\sigma$ C.L. regions, respectively.}
%\label{GRfigure1}
%\end{center}
%\end{figure}

\begin{figure}[ht]
\begin{center}
\includegraphics[width=0.45\textwidth]{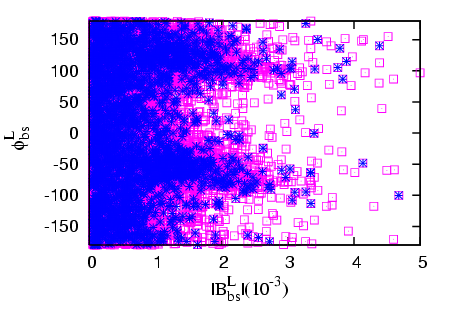}
\includegraphics[width=0.45\textwidth]{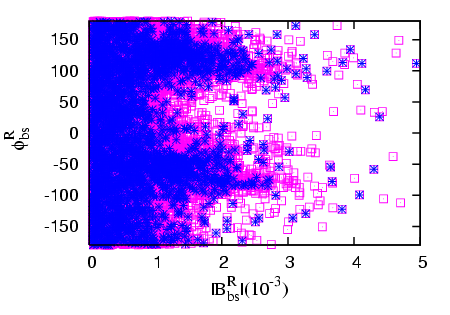}
\includegraphics[width=0.45\textwidth]{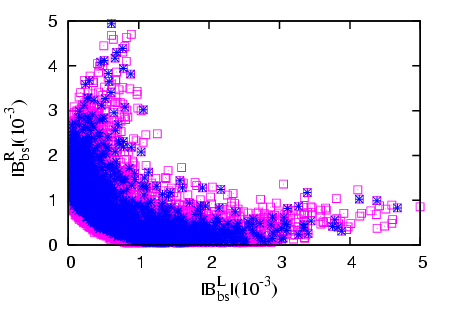}
\includegraphics[width=0.45\textwidth]{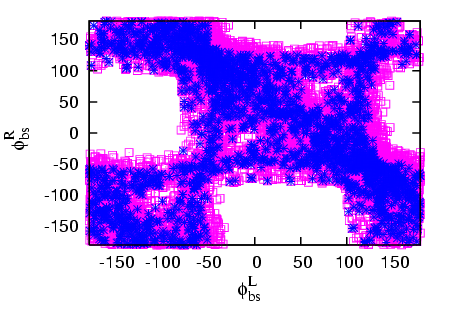}
\caption{The distributions of $|B_{bs}^{L,R}|$ and $\phi_{bs}^{L,R}$ resulting from $B_s-\bar B_s$ mixing constraints are shown. The blue and purple points can be mapped to the experimentally allowed $\{C_{B_s},\phi^{\rm NP}_{B_s}\}$ regions with $1\sigma$ and $2\sigma$ C.L., respectively. Here we did not distinct S1 and S2 solutions any more.}
\label{GRfigure1a}
\end{center}
\end{figure}

\begin{figure}[ht]
\begin{center}
\includegraphics[width=0.60\textwidth]{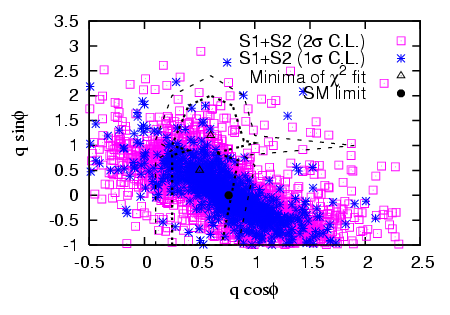}
\caption{The constraints on $B_{dd}^R$ from $q e^{i\phi}$ are illustrated. The points of $|B_{bs}^{L,R}|$ and $\phi_{bs}^{L,R}$ from Fig.~\ref{GRfigure1a} are randomly combined with the scattered points of $B_{dd}^R$ ($10^{-3}<|B_{dd}^R| < 10^{-1}$) and then mapped to the $q\cos\phi -q\sin\phi$ plane according to Eq.~(\ref{406}). The colors of the points in this plane indicate the C.L. that their inverse images represent in Fig.~\ref{GRfigure1a}.} 
%while the origin of these inverse images (i.e., with which $\phi_{B_s}^{NP}$ solution they are associated, ``S1'' or ``S2'') are not specified by different symbols any more.}
\label{GRfigure2}
\end{center}
\end{figure}

\begin{figure}[ht]
\begin{center}
\includegraphics[width=0.45\textwidth]{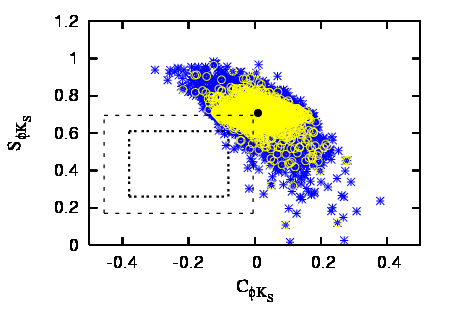}
\includegraphics[width=0.45\textwidth]{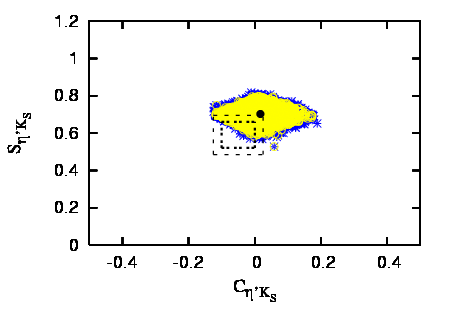}
\includegraphics[width=0.45\textwidth]{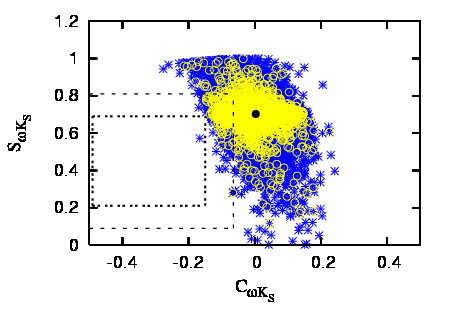}
\includegraphics[width=0.45\textwidth]{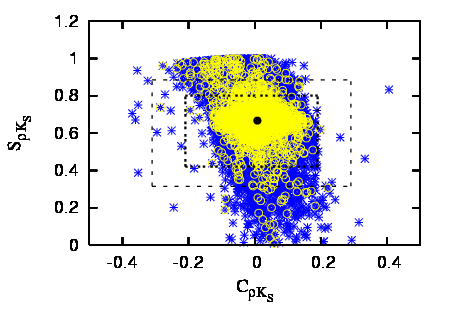}
\includegraphics[width=0.45\textwidth]{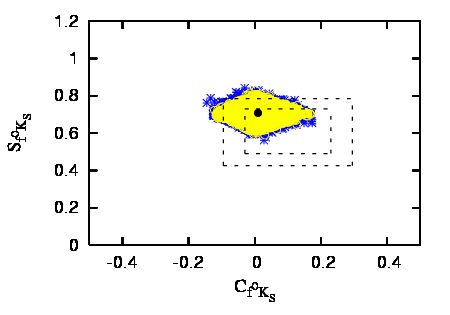}
\includegraphics[width=0.45\textwidth]{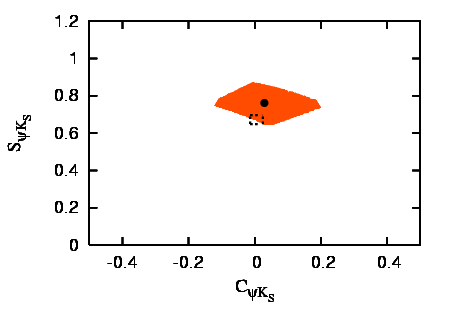}
\caption{With the values of $|B_{bs}^{L,R}|$, $\phi_{bs}^{L,R}$ and $B_{dd}^R$ fixed by $B_s-\bar B_s$ mixing and $B_d\to \pi K_S$ decay, the NP contributions to  ${\mathcal C}_{(\phi, \eta', \rho, \omega, f_0)K_S}$ and ${\mathcal S}_{(\phi, \eta', \rho, \omega, f_0)K_S}$ are illustrated in the first five panels. The colors of the points specify the C.L. that their inverse image points represent in Fig.~\ref{GRfigure1a} and Fig.~\ref{GRfigure2} (yellow denotes $1 \sigma$ C.L. in both and blue denotes $2 \sigma$ and $1.7 \sigma$ C.L., separately). In the last panel, the $CP$ asymmetries of the charmed $B_d\to \psi K_S$ decay are presented ($|V_{ub}|=3.51\times 10^{-3}$~\cite{CKMfitter08}). For each, the two boxes specify the 1$\sigma$ and $1.7\sigma$ allowed regions (except for the last panel, where only the $1 \sigma$ box is given), and the dark point denotes the SM limit.}
\label{GRfigure3}
\end{center}
\end{figure}

\begin{figure}[ht]
\begin{center}
\includegraphics[width=0.45\textwidth]{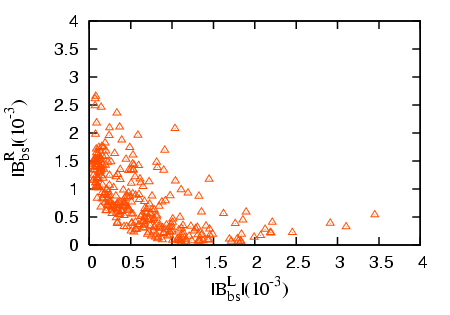}
\includegraphics[width=0.45\textwidth]{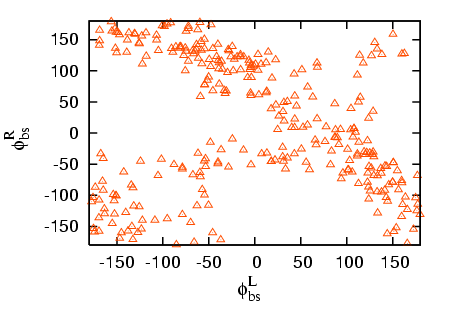}
\includegraphics[width=0.45\textwidth]{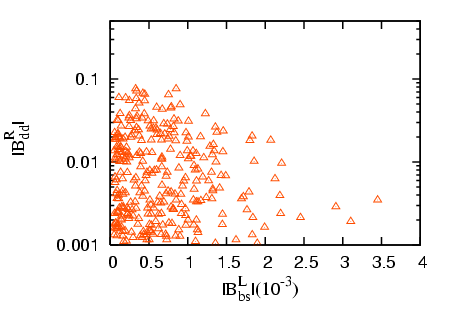}
\includegraphics[width=0.45\textwidth]{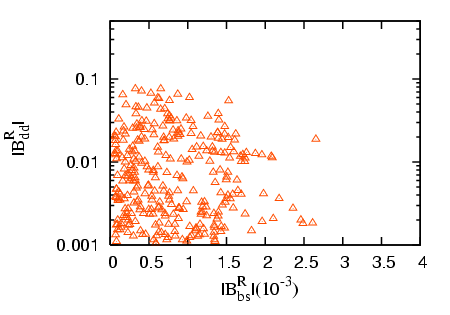}
\includegraphics[width=0.45\textwidth]{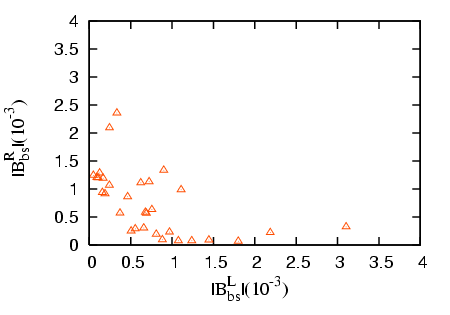}
\includegraphics[width=0.45\textwidth]{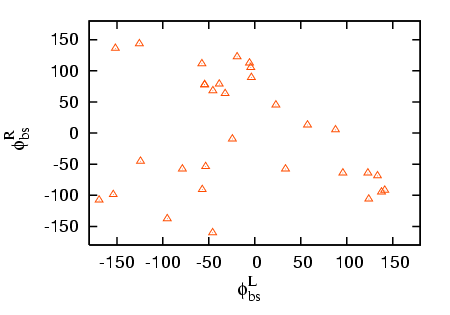}
\includegraphics[width=0.45\textwidth]{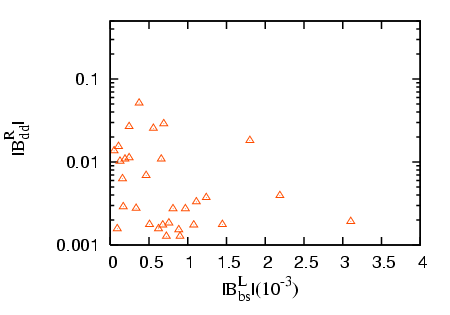}
\includegraphics[width=0.45\textwidth]{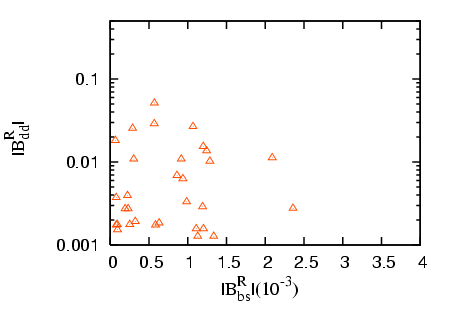}
\caption{The distributions of  $|B_{bs}^{L,R}|$, $\phi_{bs}^{L,R}$ and $B_{dd}^R$ are shown.  The values are constrained by $B_s-\bar B_s$ mixing ($2 \sigma$ C.L.) and the $\chi^2$ fit of the $B\to\pi K$ (and $B\to\pi\pi$) data ($1\sigma$ C.L.), then selected by  ${\mathcal C}_{(\phi, \eta', \rho, \omega, f_0)K_S}$, ${\mathcal S}_{(\phi, \eta', \rho, \omega, f_0)K_S}$ ($1.7 \sigma$ C.L. for the first four panels and $1.5\sigma$ C.L. for the rest).} % The non-perturbative uncertainties in the SM and NP calculations assumed to be $15\%$ and $25\%$, respectively. }
\label{GRfigure4}
\end{center}
\end{figure}

\begin{figure}[ht]
\begin{center}
\includegraphics[width=0.45\textwidth]{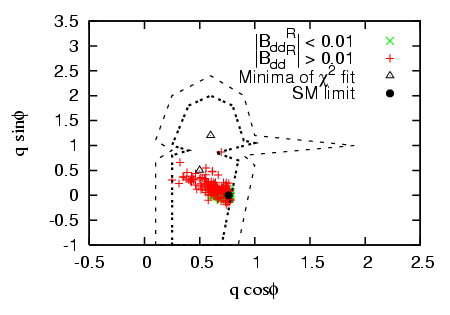}
\includegraphics[width=0.45\textwidth]{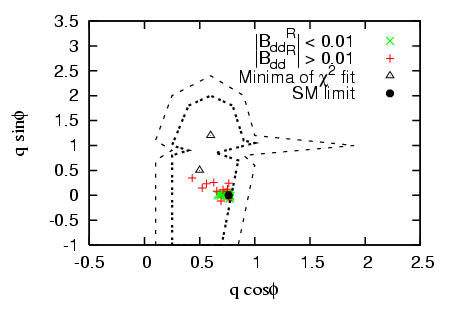}
\caption{The points in Fig.~\ref{GRfigure5} are inversely mapped to the $q\cos\phi-q\sin\phi$ plane. The parameter selection from  ${\mathcal C}_{(\phi, \eta', \rho, \omega, f_0)K_S}$, ${\mathcal S}_{(\phi, \eta', \rho, \omega, f_0)K_S}$ is shown at $1.7 \sigma$ C.L. in Fig.~\ref{GRfigure5} the left panel, and at $1.5\sigma$ C.L. in the right panel. }
\label{GRfigure5}
\end{center}
\end{figure}

As discussed in Section~\ref{section 2}, there are five free parameters in the general case: $|B_{bs}^{L,R}|$, $\phi_{bs}^{L,R}$, and $B_{dd}^R$.
Let us focus first on $B_s-\bar{B}_s$ mixing again. The general relation in Eq.~(\ref{403}) involves four of the five free parameters: $|B_{bs}^{L,R}|$ and $\phi_{bs}^{L,R}$. 
%In Fig.~\ref{GRfigure1}, we show how these parameters mediate the observables $C_{B_s}$ and $\phi_{B_s}^{NP}$, using Eq.~(\ref{403}). The physical values for $|B_{bs}^L|$ and $\phi_{bs}^L$ are then given by the points in the boxes at different confidence levels.  We illustrate their distributions  in 
In Fig.~\ref{GRfigure1a}, we show how the experimentally allowed parameter values are distributed. For each of the two top panels in Fig.~\ref{GRfigure1a}, there are two peaks and two valleys toward the right. The two peaks in the left panel correspond to the LL limit, and the two in the right panel correspond to the RR limit. Meanwhile, the points in the right panel which are associated with the LL limit  and the points in the left panel which are associated with the RR limit are localized in the regions $|B_{bs}^L| \approx 0$ and $|B_{bs}^R| \approx 0$, respectively.  As for the valleys in both panels, they correspond to the LR limit. Observe that there are two solutions for each of these three limits which are specified by a difference of $180^\circ$ either in $\phi_{bs}^L$ or $\phi_{bs}^R$ (or both). We only showed one of the two solutions in Fig.~\ref{SLfigure1}, since the difference between these two solutions can be resolved into $B_{dd}^R$ as a minus sign. 
For the two bottom panels, the three special limits LL, LR and RR correspond to the bottom boundary, the diagonal line (the one from left-bottom to right-up) and the left boundary in the left one, respectively. Clearly, for the LL (RR) limit, $|B_{bs}^L|$ ($|B_{bs}^R|$) has a relatively large value compared to the one in the LR limit, as seen in Fig.~\ref{SLfigure1}.  In the right panel, these three limits correspond to the two parallel bands $\phi_{bs}^L \sim -50^\circ, 130^\circ $, the diagonal line (the one from left-bottom to right-up) and the two parallel bands $\phi_{bs}^R \sim -50^\circ, 130^\circ $, respectively. 

In Fig.~\ref{GRfigure2},  we illustrate how the yet free parameter $B_{dd}^R$ is constrained through $q e^{i\phi}$, using the parameter values of $|B_{bs}^{L,R}|$ and $\phi_{bs}^{L,R}$ obtained in Fig.~\ref{GRfigure1a}, the relation Eq.~(\ref{406}), and the allowed range for $q e^{i\phi}$ as determined in~\cite{Fleischer:2008wb}. Compare this figure with Fig.~\ref{SLfigure2} we see that our scan selects points which
are more likely to be associated with the solution ``S1'' of $B_s - \bar B_s$ mixing. To see the reason, let us rewrite Eq. (\ref{403}) as    
%due to the $\chi^2$ fit of the $B\to\pi K$ (and $B\to\pi\pi$) data
\begin{eqnarray}
\frac{A_s^{\rm NP}}{A_s^{\rm SM}}e^{2 i \phi_s^{\rm NP}} &=& 3.59 \times 10^5  (|B_{bs}^L|^2e^{2i\phi_{bs}^L}+ |B_{bs}^R|^2e^{2i\phi_{bs}^R})  \nonumber \\
&& - 2.04 \times 10^6 |B_{bs}^LB_{bs}^R|e^{i(\phi_{bs}^L+\phi_{bs}^R)}  \label{454}
\end{eqnarray}
by using the relation Eq. (\ref{002}). Since the range of the solution ``S1'' is much larger than that of the solution ``S2'' at the same C.L. (see Table~\ref{table1}),  
 it can be understood that there are more points in the parameter space corresponding to the ``S1'' solution under the assumption of 
 flat distribution for the random values of the relevant parameters.

We illustrate the time-dependent $CP$ asymmetries of the penguin-dominated modes in Fig.~\ref{GRfigure3}, taking values of $|B_{bs}^{L,R}|$, $\phi_{bs}^{L,R}$ and $B_{dd}^R$ that are consistent with the constraints from $B_s-\bar B_s$ mixing and $B_d\to \pi K_S$ decays. As before, because of interference effects between the $B_d-\bar B_d$ mixing phase and $\phi_{bs}^L$, the points in Fig.~\ref{GRfigure3} are scattered away from the SM limits. 
%That the point number  density in Fig. (\ref{GRfigure3}) is smaller than that in Fig. (\ref{SLfigure3}) is caused by the difference of the strategies that we are taking for these two cases.
Hence, for each decay mode there are once again always some points lying in the $1\sigma$ region.  
%We universally assume a $15\%$ uncertainty in the SM calculations for each mode (as well as for the $B_d\to \pi K_S$ mode.). From the last panel where the NP effects are negligible, we see that this uncertainty is the least one necessary to simultaneously explain the experimental data of ${\mathcal C}_{\psi K_S}$, ${\mathcal S}_{\psi K_S}$ in the SM. The penguin-dominated modes have $0.5\sim 2 \sigma$ deviations for ${\mathcal C}_{f_{CP}}$, ${\mathcal S}_{f_{CP}}$, or both, except $B_d \to \rho K_S$. 
The allowed values for $|B_{bs}^{L,R}|$, $\phi_{bs}^{L,R}$ and $B_{dd}^R$ due to the correlated analysis of the $B_s-\bar B_s$ mixing and the $B_d \to (\pi, \phi, \eta', \rho, \omega, f_0)K_S$  $CP$ asymmetries are illustrated in Fig.~\ref{GRfigure4}.  In Fig~\ref{GRfigure5}, as done before, we map the points in Fig.~\ref{GRfigure4} back to the $q\cos\phi-q\sin\phi$ plane. It is straightforward to see that the points with $B_{dd}^R < -0.01$ are closer to the minima of the $\chi^2$ fit of the $B\to \pi K$ and  $B\to \pi \pi$ data, resulting in a better fit than that obtained in the SM limit.

\section{Conclusions}

In this paper, we have studied the constraints on extensions of the SM with family non-universal $U(1)'$ gauge symmetries which result from FCNC effects in the $b\to s$ transitions.   Using a  model-independent approach in which the main requirements are family universal charges for the first and second generations and small fermion mixing angles, we have performed a correlated analysis of this set of  $\Delta B =1, 2$ processes. Our results show that within this class of models, the possible anomalies in $B_s - \bar B_s$ mixing and the time-dependent  $CP$ asymmetries of the penguin-dominated $B_d \to (\pi, \phi, \eta', \rho, \omega, f_0)K_S$ decays can be accommodated in a consistent way. 

Furthermore, the constraints from $B_s - \bar B_s$ mixing may have nontrivial implications not only for the hadronic decays, but also for the leptonic or semi-leptonic decays of the $B_d$ mesons, as discussed in Section~\ref{section 2}.   As an example, recent results from BaBar~\cite{Aubert:2008ps} indicate an unexpectedly large isospin asymmetry in the low dilepton mass squared region for combined $B_d \to K l^+l^-$ and $B_d \to K^* l^+l^-$. In addition, $K^*$ longitudinal polarization and lepton forward-backward asymmetry are consistent with SM but seem to prefer a Òwrong-signÓ $C_7$ operator, suggestive of right-handed currents. We leave these interesting issues for future exploration.

\section*{Acknowledgments}

We thank Cheng-Wei Chiang and Jonathan L.~Rosner for 
useful discussions. Work at ANL is supported in part by 
the U.S. Department of Energy (DOE), Div.~of HEP, Contract
DE-AC02-06CH11357.  Work at EFI is supported in part by the 
DOE through Grant No. DE-FG02- 90ER40560.  Work at the U.~Wisconsin, Madison is supported by the DOE through Grant No. DE-FG02-95ER40896 and the Wisconsin Alumni Research Foundation.  T.L. is also supported by the Fermi-McCormick Fellowship. The work of P.L. is supported by the IBM Einstein 
Fellowship and by NSF grant PHY-0503584.

\newpage

\appendix

\renewcommand{\thesection}{Appendix \Alph{section}}

\setcounter{equation}{0}

\section{Operators of Effective Hamiltonians}

\label{Operators}

\renewcommand{\theequation}{\Alph{section}.\arabic{equation}}

A complete compilation of the
relevant operators for the $b \to s$ transitions is given in the following:

\bigskip

\leftline{\bf Current-Current Operators:}
\begin{eqnarray}
Q_{1} = \left( \bar s u \right)_{\rm V-A}
            \left( \bar u b \right)_{\rm V-A}
            &\qquad&
Q_{2} = \left( \bar s_{\alpha} u_{\beta}  \right)_{\rm V-A}
            \left( \bar u_{\beta}  b_{\alpha} \right)_{\rm V-A}
\label{eq:Q12}
\end{eqnarray}

\bigskip
\leftline{\bf QCD-Penguin Operators:}
\begin{eqnarray}
Q_{3} = \left( \bar s b \right)_{\rm V-A}
   \sum_{q} \left( \bar q q \right)_{\rm V-A}
&\qquad&
Q_{4} = \left( \bar s_{\alpha} b_{\beta}  \right)_{\rm V-A}
   \sum_{q} \left( \bar q_{\beta}  q_{\alpha} \right)_{\rm V-A}
\label{eq:Q34} \\
Q_{5} = \left( \bar s b \right)_{\rm V-A}
   \sum_{q} \left( \bar q q \right)_{\rm V+A}
&\qquad&
Q_{6} = \left( \bar s_{\alpha} b_{\beta}  \right)_{\rm V-A}
   \sum_{q} \left( \bar q_{\beta}  q_{\alpha} \right)_{\rm V+A}
\label{eq:Q56} \\
\tilde Q_{3} = \left( \bar s b \right)_{\rm V+A}
   \sum_{q} \left( \bar q q \right)_{\rm V+A}
&\qquad&
\tilde Q_{4} = \left( \bar s_{\alpha} b_{\beta}  \right)_{\rm V+A}
   \sum_{q} \left( \bar q_{\beta}  q_{\alpha} \right)_{\rm V+A}
\\
\tilde Q_{5} = \left( \bar s b \right)_{\rm V+A}
   \sum_{q} \left( \bar q q \right)_{\rm V-A}
&\qquad&
\tilde Q_{6} = \left( \bar s_{\alpha} b_{\beta}  \right)_{\rm V+A}
   \sum_{q} \left( \bar q_{\beta}  q_{\alpha} \right)_{\rm V-A}
\end{eqnarray}

\bigskip
\leftline{\bf Electroweak Penguin Operators:}
\begin{eqnarray}
Q_{7} = \frac{3}{2} \left( \bar s b \right)_{\rm V-A}
         \sum_{q} e_{q} \left( \bar q q \right)_{\rm V+A}
&\qquad&
Q_{8} = \frac{3}{2} \left( \bar s_{\alpha} b_{\beta} \right)_{\rm V-A}
         \sum_{q} e_{q} \left( \bar q_{\beta}  q_{\alpha}\right)_{\rm V+A}
\label{eq:Q78} \\
Q_{9} = \frac{3}{2} \left( \bar s b \right)_{\rm V-A}
         \sum_{q} e_{q} \left( \bar q q \right)_{\rm V-A}
&\qquad&
Q_{10} = \frac{3}{2} \left( \bar s_{\alpha} b_{\beta} \right)_{\rm V-A}
         \sum_{q} e_{q} \left( \bar q_{\beta}  q_{\alpha}\right)_{\rm V-A}
\label{eq:Q910} \\
\tilde Q_{7} = \frac{3}{2} \left( \bar s b \right)_{\rm V+A}
         \sum_{q} e_{q} \left( \bar q q \right)_{\rm V-A}
&\qquad&
\tilde Q_{8} = \frac{3}{2} \left( \bar s_{\alpha} b_{\beta} \right)_{\rm V+A}
         \sum_{q} e_{q} \left( \bar q_{\beta}  q_{\alpha}\right)_{\rm V-A}
%\label{eq:Q78} 
\\
\tilde Q_{9} = \frac{3}{2} \left( \bar s b \right)_{\rm V+A}
         \sum_{q} e_{q} \left( \bar q q \right)_{\rm V+A}
&\qquad&
\tilde Q_{10} = \frac{3}{2} \left( \bar s_{\alpha} b_{\beta} \right)_{\rm V+A}
         \sum_{q} e_{q} \left( \bar q_{\beta}  q_{\alpha}\right)_{\rm V+A}
\end{eqnarray}

\bigskip
\leftline{\bf Magnetic Penguin Operators:}
\begin{eqnarray}
Q_{7\gamma} = \frac{e}{8\pi^2} m_b \bar{s}_\alpha \sigma^{\mu\nu}
              (1+\gamma_5) b_\alpha F_{\mu\nu}
 \;\;\;\;\;\;\;\;\;\;             
%&\qquad&
Q_{8G} = \frac{g}{8\pi^2} m_b \bar{s}_\alpha \sigma^{\mu\nu}
        (1+\gamma_5)T^a_{\alpha\beta} b_\beta G^a_{\mu\nu}
\label{eq:Q78mag}
\end{eqnarray}
\bigskip

\leftline{\bf Semi-Leptonic Operators:}
\begin{eqnarray}
%Q_{7V}  = (\bar s d)_{V-A} (\bar l l)_{V} 
%&\qquad&
%Q_{7A} = (\bar s d)_{V-A} (\bar l l)_{A}
%\label{eq:Q7V7A} \\
Q_{9V}  = (\bar b s)_{V-A} (\bar l l )_{V} 
&\qquad&
Q_{10A} = (\bar b s)_{V-A} (\bar l l )_{A} \nonumber \\
\tilde Q_{9V}  = (\bar b s)_{V+A} (\bar l l )_{V} 
&\qquad&
\tilde Q_{10A} = (\bar b s)_{V+A} (\bar l l )_{A}
\end{eqnarray}

\bigskip

\leftline{\bf $B_s-\bar B_s$ Mixing Operators:}

\begin{eqnarray}
Q^{B_s}_1 = (\bar s b)_{V-A} (\bar s b)_{V-A}   &\qquad&   Q^{B_s}_2 = (\bar s_\alpha b_\beta)_{V-A} (\bar s_\beta b_\alpha)_{V-A} \nonumber \\
%  Q_3^{B_s} = (\bar{s}_R b_L) \;(\bar{s}_R b_L) &\qquad&  Q_4^{B_s} = (\bar{s}_R^\alpha b_L^\beta) \;(\bar{s}_R^\beta b_L^\alpha) \nonumber \\
%   Q_5^{B_s} = (\bar{s}_R b_L) \;(\bar{s}_L b_R) &\qquad& Q_6^{B_s} = (\bar{s}_R^\alpha b_L^\beta) \;(\bar{s}_L^\beta b_R^\alpha) 
\tilde Q^{B_s}_1 = (\bar s b)_{V+A} (\bar s b)_{V+A}   &\qquad&   \tilde Q^{B_s}_2 = (\bar s_\alpha b_\beta)_{V+A} (\bar s_\beta b_\alpha)_{V+A}   \nonumber \\
Q^{B_s}_3=\tilde Q^{B_s}_3 = (\bar s b)_{V+A} (\bar s b)_{V-A}   &\qquad&   Q^{B_s}_4=\tilde Q^{B_s}_4 = (\bar s_\alpha b_\beta)_{V+A} (\bar s_\beta b_\alpha)_{V-A} 
% \tilde  Q_3^{B_s} &=& (\bar{s}_L b_R) \;(\bar{s}_L b_R) \nonumber \\
 % \tilde Q_4^{B_s} &=& (\bar{s}_L^\alpha b_R^\beta) \;(\bar{s}_L^\beta b_R^\alpha) 
  \end{eqnarray}
where indices in color singlet currents have been suppressd for
simplicity, and $V$ and $A$ refer to $\gamma_\mu$ and $\gamma_\mu\gamma_5$, respectively.
%For the $i^{th}$ QCD and EW penguins operators, we will use $Q_i^{(q)}$ to denote the term associated with the contribution from "q" quark in this note.  

%\begin{figure}
%\\caption[]{Typical diagrams in the full theory from which the operators
%{eq:Q12}--{eq:Qnnmm} originate.  The cross in diagram (d) means
%a mass-insertion. It indicates that magnetic penguins originate from
%the mass-term on the external line in the usual QCD or QED penguin
%diagrams.}\label{fig1}
%5\begin{center}
%\vspace{0.4cm}
%\includegraphics[height=4.5 in]{oporigchiral charge}
%\end{center}
%\end{figure}

\newpage

\setcounter{equation}{0}

\section{Constraints of $ {\rm Br} (B_s \to \mu^+ \mu^-)$}

\label{Bstomumu}

\renewcommand{\theequation}{\Alph{section}.\arabic{equation}}

For an order of magnitude estimate, one can temporarily ignore the effect of renormalization group running.  Then the branching ratio of $B_s \to \mu^+ \mu^-$ is given by (also see~\cite{Barger:2009eq}) 
\begin{eqnarray}
 {\rm Br} (B_s \to \mu^+ \mu^-)
&=& \tau_{B_s} \frac{G_F^2}{4\pi} f_{B_s}^2 m_{\mu}^2 m_{B_s}
    \sqrt{1-\frac{4 m_{\mu}^2}{m_{B_s}^2}} |V_{tb}^* V_{ts}|^2 \nn \\
&& \quad \times
    \Big\{
    \left| \frac{\alpha}{2\pi \sin^2\theta_W} Y \left(\frac{m_t^2}{M_W^2}\right)  
      + 2
        \frac{B^L_{bs} B^L_{\mu\mu}}{V_{tb}^* V_{ts}}
    \right|^2   
     \nn \\
&&
     + \left| 2 
        \frac{B^L_{bs} B^R_{\mu\mu}}{V_{tb}^* V_{ts}} \right|^2 + \left| 2 
        \frac{B^R_{bs} B^L_{\mu\mu}}{V_{tb}^* V_{ts}} \right|^2+ \left| 2 
        \frac{B^R_{bs} B^R_{\mu\mu}}{V_{tb}^* V_{ts}} \right|^2
    \Big\}.
\end{eqnarray}
Here $\tau_{B_s}$ is the lifetime of $B_s$ meson, $f_{B_s}$ is the corresponding decay constant, and 
$Y(m_t^2 / M_W^2)$ in the SM part is defined in~\cite{Buchalla:1993bv}, with
\begin{eqnarray}
Y (x) = \frac{x}{8} \left ( \frac{4-x}{1-x} +\frac{3x}{(1-x)^2} \ln x \right ).
\end{eqnarray}
For $x=m_t^2 / M_W^2$, we have $Y(x) \sim 1$.
%Using the central value of the averaged
%$B_s$ lifetime $\tau_{B_s} = 1.439$ ps \cite{LEPBOSC} and $f_{B_s} = 232$ MeV,
%we obtain a SM branching ratio of $\simeq 4.1 \times 10^{-9}$.  
The present experimental exclusion limit at $2\sigma$ C.L. from a combination of CDF and D0 results is~\cite{Bernhard:2005yn} 
\begin{eqnarray}
{\rm Br}(B_s \to \mu^+\mu^-) \le 1.5 \times 10^{-7},
\end{eqnarray} 
%Additionally the projected exclusion limit, at $2\sigma$ C.L., on this process for 
%4 ${\rm fb}^{?1}$ at the Tevatron is  
%\begin{eqnarray}
%{\rm Br}(Bs \to \mu^+\mu^-) \le 2.8 \times 10^{-8}.  
%\end{eqnarray}
which thus give a constraint that
\begin{eqnarray}
    \left| 3\times 10^{-3}
      + 
        \frac{B^L_{bs} B^L_{\mu\mu}}{V_{tb}^* V_{ts}}
    \right|^2
    + \left| 
        \frac{B^L_{bs} B^R_{\mu\mu}}{V_{tb}^* V_{ts}} \right|^2  + \left|  
        \frac{B^R_{bs} B^L_{\mu\mu}}{V_{tb}^* V_{ts}} \right|^2+ \left|  
        \frac{B^R_{bs} B^R_{\mu\mu}}{V_{tb}^* V_{ts}} \right|^2      
\simlt 10^{-4} .
\end{eqnarray}
In the LR and LL limits, we have $B_{bs}^L= B_{bs}^R \sim 10^{-3}$ and $B_{bs}^L \sim 10^{-3}$, $B_{bs}^R=0$, respectively. For $(V_{l_{L,R}} \tilde \epsilon^{l_{L,R}} V_{l_{L,R}})_{22} \sim {\mathcal O}(1)$ and $g_1M_{Z'}/(g_2 M_{Z}) \sim 10-100$ (a parameter region favored by the correlated analysis of the anomalies in $B_s\to \bar B_s$ mixing and time-dependent $CP$ asymmetries of penguin-dominated $B_d$ meson decays; for details, see the last paragraph of subsection~\ref{SL}), this means that the NP contribution is comparable with the SM one. The experimental bound therefore can be easily satisfied. This conclusion also applies to the general case.

\newpage

\setcounter{equation}{0}

\section{Parameters}

\label{Parameters}

\renewcommand{\theequation}{\Alph{section}.\arabic{equation}}

The parameters used in our numerical analysis are summarized below:
\bigskip

{\bf Masses, Decay Constants, Hadronic Form Factors, and Lifetimes:}
\bigskip

%Masses, Decay Constants, Hadronic Form Factors and Lifetimes:
\begin{center}
$M_{{\pi}^{\pm}} =0.139$ GeV, \hspace{20mm} $M_{{\pi}^{0}} = 0.135$ GeV,\\
$M_{K} = 0.498$ GeV, \hspace{20mm} $M_{B} = 5.279$ GeV, \\
$M_\phi = 1.02$ GeV, \hspace{20mm} $M_\psi = 2.097$ GeV, \\
$M_{\eta^\prime} = 0.958$ GeV, \hspace{20mm} $M_{\omega} = 0.783$
GeV, \\
$M_{\rho} = 0.776$ GeV, \hspace{20mm} $M_\eta = 0.548$ GeV \\
$M_{f^0} = 0.980$ GeV, \\
$X_{\eta} = 0.57$, \hspace{20mm} $Y_{\eta} = 0.82$, \\
$m_u (\mu = 4.2 ~{\rm GeV}) = 1.86$ MeV, \hspace{20mm} $m_d (\mu = 4.2
~{\rm GeV})= 4.22$ MeV, \\ 
$m_s (\mu = 4.2 ~{\rm GeV}) = 80$ MeV, \hspace{20mm} $m_c (\mu = 4.2
~{\rm GeV}) = 0.901$ GeV, \\ 
$m_b(\mu = 4.2 ~{\rm GeV}) = 4.2$ GeV, \hspace{20mm} $m_t (\mu = M_Z)
= 171.7$ GeV,\\ 
$f_{\phi} = 237$ MeV, \hspace{20mm} $f_{B} = 190$ MeV,\\
$f_{\pi} = 130$ MeV, \hspace{20mm} $f_{K} = 160$ MeV,\\
$f_{\psi} = 410$ MeV, \hspace{20mm} $f_{\omega} = 200$ MeV, \\
$f_\rho = 209$ MeV, \hspace{20mm} $f_{f^0} = 180$ MeV,\\
$F_0^{B\pi} (0) = 0.330$, \hspace{20mm} $F_0^{BK}
(0) = 0.379$, \\ 
$F_1^{BK} (0) = 0.379$, \hspace{20mm} $A_0^{B\omega} (0) =
0.280$, \\  
$F_0^{B f} (0) = 0.250$, \hspace{20mm} $F_0^{f K} (0) = 0.030$, \\
$A_0^{B\rho} = 0.280$, \hspace{20mm} $f_{B_s} \sqrt{{\hat B}_{B_s}} =
0.262$ \\ 
$\tau_{B^0}=1.530$ ps, \hspace{20mm} $\tau_{B^-}=1.65$ ps, \\
$M_{B_s} = 5.37$ GeV, \hspace{20mm} $\tau_{B_s} = 1.47$ ps, 
\end{center}

\bigskip

{\bf QCD and EW Parameters:}
\bigskip

%QCD and EW Parameters:
\begin{center}
$G_F = 1.16639 \times 10^{-5}$ GeV$^{-2}$, 
\hspace{20mm}$\Lambda_{\overline{MS}}^{(5)} = 225$ MeV, \\
$M_W = 80.42$ GeV, \hspace{20mm} $\sin^2\theta_W = 0.23$, \\
$\eta_{2B} = 0.55$, \hspace{20mm} $J_5 = 1.627$, \\  
$\alpha_s(M_Z) = 0.118$,\hspace{20mm} $\alpha_{em} = 1/128$, \\
$\lambda = 0.2252$, \hspace{20mm} $A = 0.8117$, \\
$\bar{\rho} = 0.145$, \hspace{20mm} $\bar{\eta} = 0.339$, \\
$R_b = \sqrt{\rho^2 + \eta^2} = 0.378$. \\
\end{center}

\bigskip

{\bf Hadronic Parameters from Lattice Calculations:}
\bigskip

\begin{center}
$f_{B_s} \sqrt{{\hat B}_{B_s}} = 0.262.$
\end{center}

\newpage

%:Bibliography
\bibliographystyle{prsty}

\end{document}